\documentclass[%
 aip,
 amsmath,amssymb,
 reprint,%
]{revtex4-1}

\usepackage{graphicx}
\usepackage{dcolumn}
\usepackage{bm}

\usepackage[utf8]{inputenc}
\usepackage[T1]{fontenc}
\usepackage{mathptmx}

\newtheorem{lemma}{Lemma}[section]
\newtheorem{proposition}[lemma]{Proposition}

\begin{document}

\preprint{AIP/123-QED}

\title[]{Response and Sensitivity Using Markov Chains}

\author{Manuel Santos Guti\'errez}

\affiliation{ 
Department of Mathematics and Statistics, University of Reading, Reading, United Kingdom
}%

\affiliation{%
Centre for the Mathematics of Planet Earth, Department of Mathematics and Statistics, University of Reading, Reading, United Kingdom
}%

\author{Valerio Lucarini}%

\affiliation{ 
Department of Mathematics and Statistics, University of Reading, Reading, United Kingdom
}%

\affiliation{%
Centre for the Mathematics of Planet Earth, Department of Mathematics and Statistics, University of Reading, Reading, United Kingdom
}%

\affiliation{%
CEN-Meteorological Institute, University of Hamburg, Hamburg, Germany
}%

\date{\today}

\begin{abstract}
Dynamical systems are often subject to forcing or changes in their governing parameters and it is of interest to study how this affects their statistical properties. A prominent real-life example of this class of problems is the investigation of climate response to perturbations. In this respect, it is crucial to determine what the linear response of a system is as a quantification of sensitivity. Alongside previous work, here we use the transfer operator formalism to study the response and sensitivity of a dynamical system undergoing perturbations. By projecting the transfer operator onto a suitable finite dimensional vector space, one is able to obtain matrix representations which determine finite Markov processes. Further, using perturbation theory for Markov matrices, it is possible to determine the linear and nonlinear response of the system given a prescribed forcing. Here, we suggest a methodology which puts the scope on the evolution law of densities (the Liouville/Fokker-Planck equation), allowing to effectively calculate the sensitivity and response of two representative dynamical systems.
\end{abstract}

\maketitle

\section{Introduction}

Response theory is the scientific research area at the boundary between mathematics and physics dealing with the understanding of how complex systems react to perturbations affecting their dynamics. Even if it addresses is a very classical problem, the mathematical framework to develop such a theory is still a matter of research.

Although the construction of response theory can be approached by taking many different scientific point of views, it has been mostly driven by statistical mechanics\cite{marconireview2008}. In this context, by considering a complex system in a steady state (equilibrium or nonequilibrium) and applying some sort of forcing to the dynamics as, e.g., a change in the governing parameters, one is interested in analysing the resulting deviation from the steady state and, in some cases, its relaxation to the new one.

We can express these ideas formally as follows. Let $\{\phi^t\}_{t\geq 0}$ be a dynamical system on a compact domain $X \subset \mathbb{R}^d$ generated by the evolution equation $\dot{\mathbf{x}}=\mathbf{F}(\mathbf{x})$. We take here the continuous time point of view but one can equivalently formulate the problem for discrete dynamics. Furthermore, let $\mathcal{J}: X \longrightarrow \mathbb{R} $ denote a generic observable. Assuming that the system is in a steady state, we want to analyse the how the values of the observable $\mathcal{J}$ change when the dynamics are subject to a perturbation of the form:
\begin{equation}\label{perturbed dynamics}
    \dot{\mathbf{x}}=\mathbf{F}(\mathbf{x}) + \epsilon \mathbf{G}(\mathbf{x}),
\end{equation}
where $\epsilon \in \mathbb{R}$ is the perturbation parameter and $\mathbf{G}$ is the perturbation vector-field, which is assumed to generate (together with $\mathbf{F}$) the \emph{perturbed} dynamical system $\{\phi ^t_{\epsilon} \}_{t\geq 0}$. Thus, the value of the observable $\mathcal{J}$ will change as a result of the perturbation in the following fashion:
\begin{equation}\label{evolution of observable}
    \frac{d}{dt}\mathcal{J}(\mathbf{x})=\mathbf{F}(\mathbf{x})\cdot \nabla \mathcal{J}(\mathbf{x}) + \epsilon \mathbf{G}(\mathbf{x})\cdot \nabla \mathcal{J}(\mathbf{x}).
\end{equation}
This equation describes how the observable changes with time, but only upon integration of the perturbed system can we know what its \emph{expected} value is. The question we, thus, want to address is that of understanding and predicting the mean behaviour of quantities of interest upon modification of the governing dynamics. 

In a complex system evolving with time, the mathematical object that accounts for the statistical description of its asymptotic regime is the \emph{invariant measure}\cite{lasota}. This measure is called invariant because it does not change under the action of the system. Roughly speaking, it tells us how mass is distributed on phase-space in the far future. Thus, if a system is undergoing perturbations, its invariant measure will change and consequently, the expected value of the observables of interest. If the perturbation parameter $\epsilon$ is small enough, one can ask about the effect of the perturbation on the system at a given order of nonlinearity, this is, the response.

Let $\rho$ and $\rho _{\epsilon}$ denote the unperturbed and perturbed measures respectively. If we use the notation introduced earlier, we would like to compute the expectation value of the observable $\mathcal{J}$ in the perturbed steady state $\left\langle \rho _{\epsilon} , \mathcal{J} \right \rangle$  in terms of its former expectation value $\left\langle \rho , \mathcal{J} \right \rangle $ and suitable linear and nonlinear correction terms:
\begin{equation}\label{perturbative expansion}
\left\langle \rho _{\epsilon} , \mathcal{J} \right \rangle = \left\langle \rho  , \mathcal{J} \right \rangle + \sum _{k=1}^{\infty}\epsilon ^k \delta [\mathcal{J}]_k.
\end{equation}
 If one has access to $\delta [\mathcal{J}]_k$, one could effectively predict what the expected value in the perturbed system and describe its sensitivity to a certain perturbation.

In an isolated system in equilibrium governed by a Hamiltonian, R. Kubo\cite{kubo} found a closed set of formulas for the response describing $\delta [\mathcal{J}]_k$ and establishing the connection with the fluctuation-dissipation theorem (FDT). Such theorem can be seen as a dictionary allowing for predicting the response of a system from its free fluctuations. However, the nature of the problem tackled in Kubo's work did not allow to extend it to a more general case of nonequilibrium complex systems and the validity of the formulas was not fully addressed. In fact, deterministic dynamical systems featuring contraction of the phase-space posses invariant measures that are singular with respect to the Lebesgue measure. The absence of a regular density is a major barrier that makes impossible the straightforward application of the FDT, thus breaking the one-to-one connection between forced and free fluctuations of a system.

In the work by D. Ruelle\cite{ruelledifferentiation1997, ruelle2009}, such framework was clarified at a mathematical level, allowing to apply response theory in nonequilibrium systems. Ruelle's results are based on the use of Markov partitions and provide a rigorous response theory for Axiom A\cite{Eckmann1985} systems. Essential ingredients for the theory are the fact that the unperturbed system are structurally stable and that one can split the response operator in two parts. One refers to the contribution coming from the unstable and central manifolds and can be framed as an FDT result. The second contribution comes from the stable manifold and gives an additional term that cannot be described using the free fluctuations of the system. Hence, a suitable notion of differentiability of singular measures was established allowing to consider perturbative expansions as in Eq.~(\ref{perturbative expansion}) for a general class of deterministic dynamical systems.

The Ruelle response theory has provided a key framework for constructing algorithms aimed at practically computing the response of nonequilibrium systems to perturbations, see e.g. Ref.~\onlinecite{abramov}, and, specifically, for performing successfully climate change predictions using simple, see e.g. Ref.~\onlinecite{gritsun2017}, and comprehensive climate models, see e.g. Refs.~\onlinecite{ragone2017, lucariniclimatechange2017}.

Despite a good degree of success in the above mentioned studies, the presence of the two distinct contributions described above makes the construction of accurate response algorithms very challenging, see discussion in Ref.~\onlinecite{gritsun2017}. A different point of view, based on the study of the evolution of probabilities rather than of individual trajectories, seems necessary; see below.

\subsection{The Transfer Operator Approach}

Indeed, one can understand response theory by means of studying the effect of perturbations on the evolution of measures on phase-space as opposed to trajectories. Formulating response theory under this point of view uses essentially different machinery mostly hovering around the so called \emph{transfer operator}\cite{baladi}.

Let us suppose that $\mathbf{F}: X \subseteq \mathbb{R}^d \longrightarrow \mathbb{R}^d$ is a vector-field that generates the \emph{dynamical system} or \emph{flow} $\{\phi^t\}_{t\in \mathbb{R}}$, with $\phi ^t :X \longrightarrow X$. Then, the transfer operator semigroup $\{\mathcal{L}^{t}\}_{t\geq 0}$ can be defined as the solution of the Liouville equation\cite{lasota}:
\begin{equation}\label{liouville equation}
	\partial _t \rho (\mathbf{x},t) = -\nabla \cdot \left( \mathbf{F}\rho(\mathbf{x},t) \right),
\end{equation}
so that $\rho(\mathbf{x},t)=\mathcal{L}^t\rho _0(\mathbf{x})$ for some initial condition $\rho_0 \in L^1(X)$. In the language of probability, the transfer operator is describing the pushforward of an integrable function under the action of the dynamical system after $t$ time-units. It turns out that $\mathcal{L}^t$ is a contraction and the set $\{ \mathcal{L}^t \}_{t\geq 0}$ is a $C_0$-semigroup \cite{engelsemigroups2006}.

Eq.~(\ref{liouville equation}) symbolises a process where mass is only advected along the flow. However, in many areas and applications, the governing dynamics are stochasticly perturbed introducing uncertainty to the problem. This sort of perturbations can be modelled by stochastic processes of the form:
\begin{equation}\label{fokker-plack equation}
    \dot{\mathbf{x}} = \mathbf{F}(\mathbf{x}) + \Sigma(\mathbf{x})dW_t,
\end{equation}
where we introduce here a standard $d$-dimensional Wiener process $dW_t$ and $\Sigma (\mathbf{x}) \in \mathbb{R}^{d\times d}$ is the covariance matrix. In terms of measures, stochasticity can be translated into the fact that their evolution is not only driven by advection, but diffusion is present. The Liouville is thus transformed into the Fokker-Planck equation\cite{risken}:
\begin{widetext}
\begin{equation}
	\partial _t\rho (\mathbf{x},t)=\mathcal{A}\rho (\mathbf{x},t):=-\nabla \cdot \left(\mathbf{F}\rho(\mathbf{x},t) \right) + \frac{1}{2}\sum _{i=1}^d\sum _{j=1}^d\partial_{x_i,x_j} \Sigma \Sigma ^{\ast}(\mathbf{x})\rho (\mathbf{x},t).
\end{equation}
\end{widetext}
We have defined the differential operator $\mathcal{A}$ which can be shown to generate a $C_0$-semigroup, just as the Liouville equation does. Notice that the differential operator on the right-hand-side of Eq.~(\ref{liouville equation}) is the same as $\mathcal{A}$ if no noise is present.

The transfer operator is therefore a statistical tool rather than a dynamical one. It globally describes how densities evolve with time as opposed of giving a trajectory-wise description of the system. In fact, the spectral properties of the transfer operator carry information about the statistical properties of the system of interest. For instance, one observes that the fixed points of $\mathcal{L}^t$ are nothing else than the measure that remain fixed under the action of the dynamics, namely, the invariant measure. In fact, the ergodicity and mixing character of a dynamical system can be characterised in terms of the nature of the leading eigenvalues of $\mathcal{L}^t$ (e.g. Ref.~\onlinecite{baladi}).

Introducing perturbations on the system not only does it affect to the way trajectories evolve on phase-space but also to measures. This is reflected on the Fokker-Planck evolution equation where by considering a perturbation on the vector-field of the kind $\mathbf{F} \mapsto \mathbf{F}+\sum _k \epsilon_k \mathbf{G}_k$ we obtain a perturbed evolution law for the density:
\begin{equation}\label{perturbed fkp equation}
    \partial _t \rho (\mathbf{x},t) = \mathcal{A}\rho(\mathbf{x},t) + \sum _{k=1}^n \epsilon _k \mathcal{B}_{k}\rho (\mathbf{x},t),
\end{equation}
where $\mathcal{B}_k=-\nabla \cdot \left( \mathbf{G}_k\circ \right)$. Under some conditions\cite{engelsemigroups2006} the previous equation also generates a $C_0$-semigroup with the same functional properties as the unperturbed one. One can also introduce perturbations on the covariance matrix, which would lead to a similar equation as the one above with different perturbation operators.

\subsubsection*{Projected Transfer Operators and Markov Chains}\label{finite representation}

The transfer operator has been used to solve problems in different areas of science, see e.g. examples in geosciences Refs.~\onlinecite{tantetcrisis2018, Tantet2015,Froyland2007}. In these applications phase-space is not defined as a continuum but as finite collection of regions. As a result, each time the transfer operator is applied, a finite probability mass is shifted from one region to another one. For this reason, in applications, the transfer operator has to be understood in a finite dimensional setting.

To do this, we shall consider a finite subdivision of phase-space $X$ into $N$ regions or \emph{boxes} $\{B_i\}_{i=1}^{N}$ and define $\mathbf{1}_{B_i}$ as the characteristic function on box $B_i \subset X$. Thus, we define the projection $P_N: L^1\left(X\right) \longrightarrow U_N : = \mathrm{Span}\left(\{\mathbf{1}_{B_i}\}_{i=1}^N \right)$ as
\begin{equation}
	P_N\rho = \sum_{i=1}^N\frac{\mathbf{1}_{B_i}}{\eta \left(B_i\right)}\int _{B_i}\rho \mathbf{1}_{B_i} \eta (dx),
\end{equation}
where $\eta$ indicates some notion of volume that can depend on the problem. It follows that the operator $P_N\mathcal{L}^t: U_N \longrightarrow U_N$ admits a matrix representation:
\begin{equation}\label{projected transfer operator}
\mathcal{M}^{t}_{i,j}:=\left(P_N\mathcal{L}^t\right)_{i,j}=\frac{1}{\eta (B_i)}\int_{B_i}\mathcal{L}^t\mathbf{1}_{B_j}\eta (dx),
\end{equation}
which happens to be a Markov or \emph{stochastic} matrix. The way we practically construct $\mathcal{M} ^t$ depends on the choice of measure $\eta$. Generally, if one wants to study the asymptotic properties of the system one will take $\eta$ as the invariant measure. On the other hand, if the focus is put on the effects of the flow on the whole domain $X$, and especially if one is interested in problems of relaxation to the steady state, the correct choice would be to consider the Lebesgue measure instead\cite{tantetresonances2018}. It is worth highlighting that the properties of the dynamical system $\phi ^t$ are described by $\mathcal{L}^t$.

Perturbations on the dynamics lead to perturbations on the transfer operator $\mathcal{L}^t$ and, consequently on the matrix $\mathcal{M} ^t$. This fact motivates the problem of reverting the question, i.e., by considering perturbations of $\mathcal{M} ^t$, what can we say about the dynamics? This question has been tackled in previous work \cite{Lucarini2016, froylandoptimal2018} by suitably constructing the response operators. Alongside the development of the probabilistic theory, it was possible to establish a connection between the perturbation theory of Markov chains and linear response theory for dynamical systems, applying it to a number of low dimensional stochastic and deterministic dynamical systems.

Constructing the operator that accounts for the linear response can be a difficult and costly task\cite{abramov}. Linear response can be inferred by examining how the system of interest responds to small perturbations in the governing equations. However, this process can be an expensive procedure if one deals with high-dimensioinal models with physical relevance. Therefore, one desires to have predictive power. In this paper we demonstrate that it is possible to calculate, by using finite representations of the transfer operator, the response of a system by sampling its unperturbed dynamics and prior knowledge of the forcings applied to it. The overall goal is to provide practically usable tools for studying the response of complex nonequilibrium system, like the climate, to perturbations. In Section \ref{section perturbation theory} we will present the perturbation theory for Markovian matrices, key to construct the response operator at a coarse-grained level. In Section \ref{section link dynamical systems} we investigate how to put into practise such theory when dealing with continuous in time dynamical systems and illustrate the applicability of the methodology with two dynamical systems of interest.

\section{Perturbations of Finite Markov Chains}\label{section perturbation theory}

In this section, we consider a mixing (therefore, ergodic) Markov process with a finite number of states $N\in \mathbb{N}$. Then, a positive vector $\mathbf{u} _0\in \mathbb{R} ^N$ with $\sum _{i=1}^N\left(\mathbf{u} _0\right)_i=1$ would indicate an initial ensemble of states. The finite Markov process would be determined by a \emph{stochastic} matrix $\mathcal{M} \in \mathbb{R}^{N\times N}$ so that the sequence $\{\mathcal{M}^n \mathbf{u}_0 \}_{n=0}^{\infty}$ would be a realisation of it.

The stochastic matrix $\mathcal{M}$ enjoys some properties worth highlighting. First, we note that $\mathcal{M}_{i,j}$ is the probability of jumping into the $i$th state conditioned on being at the $j$th. Thus, it follows that
\begin{equation}
	\sum _{i=1}^N\mathcal{M}_{i,j}=1,
\end{equation}
for any $j=1,\ldots , N$. Therefore, all the entries of $\mathcal{M}$ are greater than or equal to zero. Further, since the process is mixing, it implies that there exists $p\in \mathbb{N}$, so that the entries of the matrix $\mathcal{M}^p$ are strictly greater than zero \cite{senetanonnegative1973}. Matrices satisfying this condition are called \emph{irreducible and aperiodic}\cite{baladi} or \emph{primitive}\cite{senetanonnegative1973}, although we shall refer to them as mixing by analogy with the Markov process they determine. Consequently, the Perron-Frobenius theorem\cite{senetanonnegative1973} holds for this matrix, meaning that there exists a leading eigenvalue $\lambda _1$ with positive eigenvector $\mathbf{u}$. By virtue of the spectral properties of stochastic matrices, it turns out that $\lambda _1 = 1$ and $\mathbf{u}$ solves
\begin{equation}\label{eigenvalue problem}
	\mathcal{M}\mathbf{u}=\mathbf{u}.
	\end{equation}
This means that $\mathbf{u}$ remains invariant under the action of $\mathcal{M}$. If, in addition, $\mathbf{u}$ is normalised so that its entries sum up to one, we call $\mathbf{u}$ the\emph{ invariant measure} of the process and it follows that $\lim _{p\rightarrow\infty}\mathcal{M}^p \mathbf{u}_0=\mathbf{u}$, for any normalised non-zero vector $\mathbf{u} _0 \in \mathbb{R} ^N$.

The next step now is to perturb the stochastic matrix and express the resulting \emph{perturbed invariant measure} as a perturbative expansion. In what follows, we generalise what reported in Ref.~\onlinecite{Lucarini2016}. For that, we consider a perturbation of $\mathcal{M}$ of the form:
\begin{equation}\label{perturbation}
	\mathcal{M} \longrightarrow \mathcal{M} + \sum_{i=1}^n\epsilon_k m_k,
\end{equation}
where $\epsilon_1,\ldots ,\epsilon _n \in \mathbb{R}$ and $m_1,\ldots ,m_n\in \mathbb{R}^{N\times N}$. The matrices $m_k$ are what we will call the \emph{perturbation matrices}.  Note that the perturbed matrix $\mathcal{M} +  \sum_{k=1}^n\epsilon_k m_k$, must also be a stochastic matrix if we want it to describe a Markov process. This requirement corresponds to having
\begin{equation}\label{c1}
	\sum _{i=1}^N\left(m_k\right)_{i,j}=0,
\end{equation}
for any $k$ and $j$. This assures that the columns of $\mathcal{M} +  \sum_{k=1}^n\epsilon_k m_k$ add up to one. More, non-negativity must be preserved, so not all choices of $\epsilon _k$ are valid. For this, we define
\begin{equation}
	\epsilon^-=\min \{ \epsilon \in \mathbb{R} : \forall i, j \in \{1,\ldots, N\}, \mathcal{M} _{i,j} +\epsilon\sum _{k=1}^n\left(m_k\right)_{i,j} \geq 0  \},
\end{equation}
and
\begin{equation}
	\epsilon^+=\max \{ \epsilon \in \mathbb{R} : \forall i, j \in \{1,\ldots, N\}, \mathcal{M} _{i,j} +\epsilon\sum_{k=1}^n\left(m_k\right)_{i,j} \geq 0  \}.
\end{equation}
Hence, to ensure non-negativity of the perturbed Markov process, we must have that $\max_k \epsilon _k \in \left[\epsilon^- , \epsilon^+ \right]$. To ensure that the latter interval is non-empty, we need to make sure that $\epsilon ^- < 0$ and $\epsilon ^+ >0$. Suppose that $\mathcal{M}_{i_1,j_1}=\mathcal{M}_{i_2,j_2}=0$ and $\left(\sum _{k}m_k \right)_{i_1,j_1},\left(\sum _{k}m_k \right)_{i_2,j_2}<0$. This would imply that $\epsilon ^-=\epsilon ^+=0.$ In this case, we have non-admissible perturbations.

Again, by virtue of the Perron-Frobenius theorem, such perturbed matrix will have a dominant eigenvalue, whose value is $1$ and its associated eigenvector $\mathbf{v}=\mathbf{v}(\epsilon_1,\ldots ,\epsilon _n)$ is strictly positive. In other words, $\mathbf{v}$ solves:
\begin{equation}\label{perturbed eigenvalue problem}
	\left( \mathcal{M} +  \sum_{k=1}^n\epsilon_k m_k \right)\mathbf{v}= \mathbf{v}.
\end{equation}
The goal is to express the perturbed invariant measure $\mathbf{v}$ in terms of $\mathcal{M}, m_1,\ldots , m_n, \epsilon_1,\ldots$ and $\epsilon _n$. Not only do we want to calculate $\mathbf{v}$ but also describe how the unperturbed measure $\mathbf{u}$ responds at a given power of $\epsilon _k$.

Using multiindex notation, we suppose a formal expansion in powers of $\epsilon_1, \ldots , \epsilon _n$: 
\begin{equation}
\mathbf{v} = \mathbf{u} +\sum_{|\alpha|=1}^{\infty}(\epsilon _1,\ldots,\epsilon _n)^{\alpha} \mathbf{w} _{\alpha},
\end{equation}
 where $\mathbf{w}_{\alpha}=\frac{1}{\alpha !}\left(\partial _{\epsilon_1},\ldots,\partial _{\epsilon _n}\right)^{\alpha}$. Substituting this expression in Eq.~(\ref{perturbed eigenvalue problem}) we obtain:
 \begin{widetext}
\begin{equation}\label{algebraic expansion}
\left(\mathcal{M} + \sum _{k=1}^{n}\epsilon_k m_k \right) \left(\mathbf{u} +\sum_{|\alpha|=1}^{\infty}(\epsilon _1,\ldots,\epsilon _n)^{\alpha} \mathbf{w} _{\alpha}\right) = \mathbf{u} +\sum_{|\alpha|=1}^{\infty}(\epsilon _1,\ldots,\epsilon _n)^{\alpha} \mathbf{w} _{\alpha}.
\end{equation}
\end{widetext}
Gathering the terms for $| \alpha | =1$ we get:
\begin{align*}
	\mathcal{O}(\epsilon _k): \left( 1 - \mathcal{M} \right)\partial _{\epsilon _k}\mathbf{v} = m_k \mathbf{u},
\end{align*}
for $k=1,\ldots,n$. The matrix $1-\mathcal{M}$ cannot be inverted since $1$ is an eigenvalue of $\mathcal{M}$; we shall discuss this issue in the next section. At the moment, we directly apply the inverted matrix to find that
\begin{equation}\label{linear response operator}
\partial _{\epsilon _k}\mathbf{v} = 	\left( 1 - \mathcal{M} \right)^{-1}m_k \mathbf{u}.
\end{equation}
We define $\Psi _k = \left( 1- \mathcal{M} \right)^{-1}m_k$ as \emph{linear response operator}. This matrix has also been named a \emph{differential} matrix\cite{schweitzerperturbation1968}. If we repeat the process for the second order terms ($|\alpha | =2$) we get:
\begin{align*}
	\mathcal{O}\left(\epsilon_k^2\right): \frac{1}{2!}\partial^2_{\epsilon _k}\mathbf{v} &= 	\left( 1 - \mathcal{M} \right)^{-1}m_k \partial _{\epsilon _k}\mathbf{v} = \left( \Psi_k \right)^2\mathbf{u} \\ 	\mathcal{O}\left(\epsilon_k\epsilon _l\right): \partial^2_{\epsilon _k,\epsilon_l}\mathbf{v} &= 	\left( 1 - \mathcal{M} \right)^{-1}m_l \partial _{\epsilon _k}\mathbf{v} + \left( 1 - \mathcal{M} \right)^{-1}m_k \partial _{\epsilon _l}\mathbf{v} \\&= \Psi_k \Psi_l \mathbf{u}  +\Psi_l \Psi_k \mathbf{u}.
\end{align*}
Thus, we inductively construct the whole expansion as:
\begin{align}\label{algebraic expansion 1}
	\mathbf{v} &= \mathbf{u} + \sum_{i=1}^{\infty} \left(\sum _{k=1}^{n}\epsilon _k \Psi_k \right)^i\mathbf{u}  =\left(1 - \sum _{k=1}^n\epsilon _k \Psi_k \right)^{-1}\mathbf{u}.
\end{align}
This formula provides a tool to predict the perturbed invariant measure for as long as it converges. Perturbative formulas like this were found in the probability literature\cite{schweitzerperturbation1968} and later revisited when linking them to the study of the response of dynamical systems\cite{Lucarini2016, froylandoptimal2018}.

Moreover, by simply considering the adjoint matrices, it is possible to develop a response theory for observables. Indeed, let $\mathcal{J}\in \mathbb{R}^{N}$ represent a generic coarse-grained observable. Then, its expectation value with respect to the perturbed stationary vector $\mathbf{v}$ satisfies:
\begin{widetext}
\begin{equation}
\left\langle \mathcal{J} , \mathbf{v} \right\rangle = \left\langle \mathcal{J} ,\mathbf{u} + \sum_{i=1}^{\infty} \left(\sum _{k=1}^n\epsilon _k \Psi_k \right)^i\mathbf{u}  \right\rangle   =\left\langle \left( 1 + \sum_{i=1}^{\infty} \left(\sum _{k=1}^n\epsilon _k \Psi_k \right)^i\right)^{\top}\mathcal{J} ,\mathbf{u}  \right\rangle =\left\langle  1 + \sum_{i=1}^{\infty} \left(\sum _{k=1}^n\epsilon _k \Psi_k^{\top} \right)^i \mathcal{J} ,\mathbf{u}  \right\rangle, \label{algebraic expansion 2}
\end{equation}
\end{widetext}
where $\left\langle \cdot , \cdot \right\rangle$ denotes the pairing between measures and observables. The advantage of formulas (\ref{algebraic expansion 1}) and (\ref{algebraic expansion 2}) is that they allow us to identify the response to perturbations at an arbitrary order of nonlinearity including the linear case, which has a special relevance in the physical literature. As a consequence, if one is interested in calculating the sensitivity of the expectation value of some observable $\mathcal{J}$ with respect to changes caused by $\epsilon _k m_k$, one has to truncate the series expansion in the first order term. However, these formulas are still purely formal. We need a deeper understanding of what we mean with $\left(1-\mathcal{M} \right)^{-1}$ and for what values of $\epsilon _k$ they are useful.

\subsection{Well-posedness and Invertibility of $1-\mathcal{M}$}

In this section we will clarify the framework for which the response formulas presented above work. First of all, we will revisit the problem of the non-invertibility of $1-\mathcal{M}$. Secondly, we will assess the convergence of the power expansion in Eq.~(\ref{algebraic expansion 1}) and include a comment on the stability of the leading eigenvalues of $\mathcal{M}$.

The linear response operator is not well defined \emph{a priori} as $1$ is an eigenvalue of the matrix $\mathcal{M}$  and hence $1-\mathcal{M}$ is not invertible.  However, we can define a more suitable normed space for which the norm of $ \mathcal{M}$ is less than one, making $1-\mathcal{M}$ invertible. The idea relies on the fact that the vector space $\mathbb{R}^{N}$ on which $\mathcal{M}$ is defined admits a splitting of the form: 
\begin{equation}\label{decomposition}
	\mathbb{R}^{N}=\mathrm{Span}\left( \mathbf{u} \right) \oplus V,
\end{equation}
where $V$ is the invariant subspace generated by the generalised eigenvectors of $\mathcal{M}$ associated with the eigenvalues distinct to $1$ and $\mathrm{Span}\left( \mathbf{u} \right)$ is the vector subspace spanned by $\mathbf{u}$. This space can also be regarded as the kernel of the functional $\iota:\mathbb{R}^N\longrightarrow \mathbb{R}$ given by $\iota\left( \mathbf{x} \right)=\sum _i(\mathbf{x})_i$. Indeed, one observes that if $\mathbf{u}_k$ is a generalised eigenvector of $\mathcal{M}$, the identity $\left( \mathcal{M} - \lambda_k \right)^p\mathbf{u}_k=\mathbf{0}$ holds for some $p\in \mathbb{N}$. Hence,
\begin{align*}
    \iota\left( \left( \mathcal{M} - \lambda_k \right)^p\mathbf{u}_k \right)&= \sum _{j=0}^p \begin{pmatrix}p \\ j\end{pmatrix}\lambda^j _{k}\iota \left(\mathcal{M}^{p-j}\mathbf{u}_k\right)\\&=\left(1-\lambda^p_k \right)\iota(\mathbf{u}_k) = 0.
\end{align*}
This implies that $\iota (\mathbf{u}_k)=0$, by virtue of the mixing hypothesis. Furhtermore, noting that $\iota \left( m_k \mathbf{x} \right)=0$ for any $\mathbf{x}\in \mathbb{R}^N$, we can summarise the idea in the following statement:
\begin{proposition}\label{proposition other}
	Let $\mathcal{M}\in \mathbb{R}^{N\times N}$ be a mixing stochastic matrix with invariant measure $\mathbf{u}$. Then, for any $\mathbf{x}\in \mathbb{R} ^N$, $m_k\mathbf{x}\in V$ and $1-\mathcal{M}$ is invertible on $V$.
\end{proposition}

Recall that the linear response operator in Eq.~(\ref{linear response operator}) only requires the evaluation of $\left(  1 - \mathcal{M}\right)^{-1}$ after having calculated $m_k \mathbf{u}$, so writing the inverse explicitly is not that big an abuse. However, we must underline that, numerically, we cannot directly invert $1-\mathcal{M}$. To overcome this problem, we must deflate the matrix, removing the dependence on the dominant eigenspace. Concretely, we have to consider the \emph{group inverse}\cite{Funderlic1986} $\mathcal{Z}$ of a Markov chain, which is defined as follows:
\begin{equation}
	\mathcal{Z}=\left( 1 - \mathcal{M} + \mathcal{M} ^{\infty} \right)^{-1},
\end{equation}
where the matrix $\mathcal{M} ^{\infty}$ is defined as $\lim _{p\rightarrow \infty}\mathcal{M} ^p$. This matrix can be shown to be equal to the matrix whose columns are all equal to  the invariant measure $\mathbf{u}$. Intuitively, $\mathcal{M} ^{\infty}$ can be seen as a rank-one projector onto $\mathrm{span}\left( \mathbf{u} \right)$. In fact, $1$ is not an eigenvalue of $\mathcal{M} - \mathcal{M}^{\infty}$, making $1 - \mathcal{M} + \mathcal{M} ^{\infty}$ invertible. The linear response operator is, thus, given by
\begin{equation}
	\partial_{\epsilon_k}\mathbf{v}=\Psi _k\mathbf{u}=\left( 1 - \mathcal{M} \right)^{-1}m_k\mathbf{u} = \mathcal{Z}m_k\mathbf{u}.
\end{equation}
This way we find a practical way of computing the linear response operator.

To assess convergence, we want to be certain that the $L^1$ norm of the series $\sum _{k=1}^{\infty}\left(\epsilon _1\Psi _1 + \ldots +\epsilon_n\Psi _n \right)^k$ does not go to infinity. For this problem we introduce the matrix norm $\|\cdot  \|_{1^{\ast}}$ which we define as the norm $\| \cdot \| _1$ restricted to $V$. We are now in conditions of applying the ratio test in Eq.~(\ref{algebraic expansion 1}):
\begin{widetext}
\begin{align*}\label{expandise fefa}
	\frac{\left \| \left( \sum_k\epsilon _k\Psi_k \right)^{i+1}\mathbf{u}\right\|_1}{\left \| \left(\sum_k\epsilon _k\Psi_k \right)^{i}\mathbf{u}\right\|_1} &=  \frac{\left\| \left(\sum_k\epsilon _k\Psi_k\right)\left(\sum_k\epsilon _k\Psi_k \right)^i\mathbf{u}\right\|_1}{\left\|\left(\sum_k\epsilon _k\Psi_k \right)^i\mathbf{u}\right\|_1} \leq \left\| \sum_{k=1}^n\epsilon _k\Psi_k \right\| _1 = \left \| \left( 1 - \mathcal{M} \right)^{-1} \sum _{k=1}^n \epsilon _k m_k  \right\|_1 \leq \left \|  \left( 1 - \mathcal{M} \right)^{-1} \right \|_{1^{\ast}} \left\| \sum_{k=1}^n \epsilon _km_k \right\|_1  \\&\leq  \left(  1 - \| \mathcal{M} \|_{1^{\ast}}  \right)^{-1} \left \|  \sum _{k=1}^n\epsilon _km_k \right \|_1 \leq  \left(  1 - \| \mathcal{M} \|_{1^{\ast}} \right)^{-1} \max_k\{\epsilon_k\} n \max_k\{ \|  m_k  \|_1\}.
\end{align*}
\end{widetext}
Since we want that the ratio remains lower than $1$ to ensure (absolute) convergence, we choose $\epsilon_1,\ldots, \epsilon _n$ so that
\begin{equation}\label{size epsilon}
	\vert  \max_k\{  \epsilon _k \} \vert <\epsilon _{max} :=\frac{1 - \| \mathcal{M} \|_{1^{\ast}}}{ n \max_k\{ \|  m_k  \|_1\}}.
\end{equation}
By referring to the mixing character of $\mathcal{M}$ we conclude that $\epsilon _{max}$ is finite and positive. Hence, $\epsilon_{max}$ establishes a tolerance on the size of the perturbation. We would like to underline that the condition for these response formulas to work can be translated to the fact that one requires that the system has to mix mass sufficiently quickly. 

The number $\left \| \mathcal{M} \right\|_{1^{\ast}}$ is known as the \emph{ergodicity coefficient}\cite{senetacoefficients1979} and  can serve as a condition number for the Markov chain in the sense that this quantity serves to estimate the norm of the difference between a perturbed and the unperturbed invariant measures\cite{senetasensitivity1993}. See Ref.~\onlinecite{inubushiunpredictability2019} for a practical use of the ergodicity coefficient. However, the ergodicity coefficient does not tell us how perturbations affect the localisation of the eigenvalues of the matrix $\mathcal{M}$, something crucial if one wants to control the spectral gap. This problem is tackled by means of analysing the stability of the leading non-unit eigenvalues and summarised in the following result found in Ref.~\onlinecite{mitrophanovstability2003}:

\begin{proposition}
   Let $\mathcal{M}\in \mathbb{R} ^{N\times N}$ be a diagonalisable, irreducible and aperiodic stochastic matrix. Let $X\in \mathbb{R} ^{N\times N}$ be a non-singular matrix such that $\mathcal{M}=X^{-1}\Lambda X$ with $\Lambda$ being the diagonal matrix containing the eigenvalues $1,\lambda_2,\ldots ,\ldots \lambda_N$, of $\mathcal{M}$. Suppose that
    \begin{equation}\label{mitrophanov 2003}
         \kappa (X)\max_k\{|\epsilon _k|\}n\max_k\{\|m_k\|\} < \frac{1}{2}\min_{1\leq j \leq N}\{ 1 - | \lambda _j|\}.
    \end{equation}
    Then, the perturbed chain $\mathcal{M}+\sum_{k=1}^n\epsilon _km_k$ has a unique invariant measure.
\end{proposition}
The proof of this result relies on classical perturbation theory, in particular on the Bauer-Fike theorem\cite{bauerfike}. With a bit more work, one can deduce a bound on the rate of convergence to equilibrium of the perturbed chain, also presented in the work cited above.

The condition shown in Eq.~(\ref{mitrophanov 2003}) can very restrictive if the process is governed by a highly non-normal Markov matrix, as in this case $\kappa \left(X\right)$ can be very large. However, in order to preserve the spectral gap, the only thing needed is a good conditioning of the eigenvalues closest to the unit circle. Hence, in order to find a sharper stability bound like in Eq.~(\ref{mitrophanov 2003}) the \emph{eigenvalue condition number}\cite{wilkinson} might be the object to look at.

\section{Link to Continuous Time Dynamical Systems}\label{section link dynamical systems}

In this section, we will investigate how to obtain Markov chains from continuous time dynamical systems up to finite precision via focusing on the evolution of densities given by the Fokker-Planck equation. The ultimate target will be to apply the theory for Markov chains presented earlier to asses the response to perturbations of two simple dynamical systems featuring different characteristics. 

In general, we proceed the following way. Let $dt>0$, and let us express the Fokker-Planck equation (\ref{fokker-plack equation}) to first order as:
\begin{equation}\label{discrete fokker-plank euqation}
    \rho(\mathbf{x},t+dt) =  \rho(\mathbf{x},t) + dt \mathcal{A} \rho (\mathbf{x},t) + \mathcal{O}(dt^2).
\end{equation}
The idea is to view the right hand side of the previous equation as the pushforward of the measure $\rho(\mathbf{x},t)$ to $\rho(\mathbf{x},t+dt)$, namely, $\mathcal{L}^{dt}\rho(\mathbf{x},t)$. Therefore, when considering the projection of $\mathcal{L}^{dt}$ we would be introducing Markov chains, as clarified earlier.

Furthermore, a perturbation of the driving vector-field $\mathbf{F}\rightarrow \mathbf{F}+\sum _k \epsilon_k \mathbf{G}_k$ would incur a modification on the Fokker-Planck equation. Again, to first order:
\begin{equation}
    \rho(\mathbf{x},t+dt) = \rho(\mathbf{x},t) + dt\mathcal{A} + dt\sum _k \epsilon _k \mathcal{B}_k + \mathcal{O}(dt^2).
\end{equation}
It is natural then to regard $\mathcal{B}_k$ as the perturbation operators. In what follows we will frame this perturbation problem in a finite-precision setting by means of the finite representation of the transfer operator.

\subsection{Transition Matrices}

We observed that matrices introduced in Eq.~(\ref{projected transfer operator}) can be understood as Markov processes describing the probability of mass transitioning between regions of phase-space. We also noted that the measure $\eta$ employed in the projection of the transfer operator determines the algorithms used. Generally, in dynamical systems, one is interested in the properties of the system in its asymptotic regime. This means that the projection of the transfer operator in this case has to be done with respect to the invariant measure. 

In order to sample the invariant measure, long integrations of the system are needed to explore the region of phase-space where the long-term dynamics occur. Due to finite precision, integrations can be translated into time-series with equal time-step $dt$ creating a cloud of points on the domain. After transients have died, sample points will populate certain region on phase-space, possibly \emph{shadowing}\cite{robinson} the dynamics on an attractor. Such region can be subdivided into $N$ boxes $\{B_i\}_{i=1}^N$ with Lebesgue-zero measure intersections. If $\mathcal{S}$ denotes all the sample points living in $\cup_i B_i$, the matrix $\mathcal{M}$ in Eq.~(\ref{projected transfer operator}) is constructed as:
\begin{equation}\label{sample points transition matrix}
    \mathcal{M}^{dt}_{i,j}=\frac{\# \{ \mathcal{S} \cap B_i \cap \phi ^{-dt}B_j \}}{\#\{\mathcal{S}\cap B_j\}},
\end{equation}
where $\# $ is the counting measure. Essentially, this formula is counting the transitions from box to box that sample points of the time-series do after a lag of $dt$ time-units. By construction, it immediately follows that $\mathcal{M}^{dt}$ is a stochastic matrix. A matrix constructed this way is called a \emph{transition matrix}.

This representation of the transfer operator is what we are going to use to approximate the right-hand-side of Eq.~(\ref{discrete fokker-plank euqation}) to first order. If instead one is considering the perturbed problem the Fokker-Planck equation is modified (see Eq.~(\ref{perturbed fkp equation})), and a suitable matrix approximation of the perturbation operator $\mathcal{B}_k=-\nabla \cdot \left( \mathbf{G}_k\circ \right)$ is needed. Given the discrete setting, we can use finite-volume schemes to approximate the differential operator $\mathcal{B}_k$. Of course, the way we do this depends on the particular problem. In the next sections we will put this methodology into practise with two examples.

\subsection{A Stochastic Dynamical System}


To illustrate the applicability of the methodology described above, we consider the Ornstein-Uhlenbeck (O-U) process as a test case. The O-U process in $\mathbb{R}^d$ is a stochastic process $\{X(t)\}_{t\geq 0}$ of the form:
\begin{equation}\label{ou process 1}
    dX = AXdt + \Sigma dW_t,
\end{equation}
were $A \in \mathbb{R}^{d\times d}$ and models the linear and deterministic component of the process. The stochastic part of the process is given by $\Sigma dW_t$, where $\Sigma\in \mathbb{R}^{d\times d}$ indicates the correlations in the $d$-dimensional Wiener process $dW_t$. In order for this process to posses an invariant measure, the matrix $A$ is required to have eigenvalues with strictly negative real part\cite{daprato1996}. If this condition is met, the process will be stable and the statistics converge to a normal distribution.

The O-U process can also be viewed from the statistical scope by considering its associated Fokker-Planck equation:
\begin{align*}
    \partial _t \rho (\mathbf{x},t)=& - \sum_{k=1}^2\partial _{x_k}\left(\sum _{l=1}^2 A_{k,l}x_l\rho(\mathbf{x},t) \right) \\&+ \frac{1}{2}\sum _{k=1}^2\sum_{l=1}^2 \partial ^2_{x_k,x_l}\left( \left(\Sigma \Sigma ^{\ast}\right)_{k,l}\rho(\mathbf{x},t) \right),
\end{align*}
where $\rho \in L^1(X)$ is a density for each value of time $t$.

We consider the two dimensional O-U process given by:
\begin{equation}
    A=\begin{bmatrix}  1 & 0 \\ 0 & 1 \end{bmatrix}, \text{ and } \Sigma=\begin{bmatrix}  1 & 0 \\ 0 & 1 \end{bmatrix}.
\end{equation}
The variables in this process are totally uncorrelated and we shall investigate the response of the system when its mean is shifted and correlations are introduced in the noise. Thus, the process we are going to study is given by:
\begin{equation}\label{ou process 2}
    dX^{\epsilon _1 , \epsilon _2} = A\left(X^{\epsilon_1,\epsilon _2}-\epsilon_1 \mu \right)dt + \sqrt{\Sigma \Sigma ^{\ast} + \epsilon _2 E}dW_t,
\end{equation}
where
\begin{equation}\label{perturbation ou}
    \mu=\begin{bmatrix}  1 \\ 0 \end{bmatrix}, \text{ and } E=\begin{bmatrix}  0 & 1 \\ 1 & 0 \end{bmatrix},
\end{equation}
and $\epsilon_1, \epsilon_2 \in \mathbb{R}$ and have to satisfy the conditions for the perturbed process to be well defined, namely, $\Sigma \Sigma ^{\ast} + \epsilon _2 E$ has to be symmetric positive definite. The modified Fokker-Planck equation therefore is,
\begin{align*}
    \partial _t \rho (\mathbf{x},t)=& - \sum_{k=1}^2\partial _{x_k}\left(\sum _{l=1}^2 A_{k,l}x_l\rho(\mathbf{x},t) \right) \\&+ \frac{1}{2}\sum _{k=1}^2\sum_{l=1}^2 \partial ^2_{x_k,x_l}\left( \left(\Sigma \Sigma ^{\ast}\right)_{k,l}\rho(\mathbf{x},t). \right) \\ &-\epsilon _1 \sum_{k=1}^2\partial _{x_k}\left(\sum _{l=1}^2 \left(A\mu\right)_l\rho(\mathbf{x},t) \right)\\ & + \frac{\epsilon _2}{2}\sum _{k=1}^2\sum_{l=1}^2 \partial ^2_{x_k,x_l}\left( \left(E\right)_{k,l}\rho(\mathbf{x},t) \right).
\end{align*}
In a concise manner, it can be written as
\begin{equation}\label{perturbed abbreviated fokker planck}
    \partial _t \rho (\mathbf{x},t) = \mathcal{A}\rho (\mathbf{x},t) + \epsilon _1 \mathcal{B}_1\rho(\mathbf{x},t) + \epsilon_2\mathcal{B}_2\rho(\mathbf{x},t),
\end{equation}
where the differential operators $\mathcal{B}_1$ and $\mathcal{B}_2$ are defined by analogy.

As observed in the previous section, by differencing the time-variable with a time-step of $dt>0$, we immediately obtain a discrete time evolution of densities to first order. Thus, the right-hand-side of the evolution equation is pushing forward the measures at time $t$ to resulting measures at time $t + dt$. Recall that this is what the transfer operator $\mathcal{L}^{dt}$ precisely does.

\subsubsection*{Numerics: transition and perturbation matrices}

In order to construct a transition matrix, one has to have in hand a compact region of phase-space where all the long-term dynamics occur. Unfortunately, in the case of the O-U process, phase-space is unbounded because of the presence of white noise. However, one can practically restrict the phase-space to a compact rectangle that with high probability encloses the whole integration of the process.

Let $\{B_i\}_{i=1}^{2^N}$ be a collection of boxes covering four standard deviations of the process on each direction. In the experiments performed we have considered $2^{N}$ boxes by means of discretising each axis in $2^{N/2}$ equally sized segments. We have examined the values $N=10,12$ and $14$ trying to keep a balance between numerical tractability and precision. The box subdivision of phase-space is done using the \textsc{Matlab} package GAIO\cite{gaio}. We then construct the transition matrix $\mathcal{M}^{dt}$ as in Eq.~(\ref{sample points transition matrix}), where $\phi^{\circ}$ is now replaced by the process $X(\circ)$. To obtain the long time-series, we integrate it using an Euler-Maruyama scheme for $10^{6}$ time-units with a time-step of $dt=10^{-2}$ time-units. $X \left({-dt}\right)B_j$ denotes the set of points on phase-space that will end up in $B_j$ after waiting $dt$ time-units.

Regarding the perturbation operators present in Eq.~(\ref{perturbed abbreviated fokker planck}), we examine how two implement them so that they are compatible with the domain discretisation carried out for the transition matrices. Suppose that the box $B_i$ has center $c^i_{k,l}$ so that $c^i_{k \pm 1,l}=c^i_{k,l} \pm [\delta _{x_1} , 0]$, where $\delta _{x_1}$ is the distance between consecutive boxes along the $x_1$-direction (the same for the $x_2$-direction). Then, the derivative of $\rho$ with respect to $x_1$ at $c^i_{k,l}$ is given by
\begin{equation}\label{centre difference}
\partial _{x_1} \rho(c^i_{k,l}) = \frac{\rho(c^i_{k+1,l})-\rho (c^i_{k-1,l})}{2\delta_{x_1}} + \mathcal{O}\left(\delta _{x_1} ^2\right).
\end{equation}
The same scheme is used for the $x_2$-direction. For the second and cross derivatives, we implemented the usual second order discretisation:

\begin{widetext}
\begin{align}
  	\partial^2 _{x_1} \rho(c^i_{k,l})&=\frac{\rho(c^i_{k+1,l}) - 2\rho  (c^i_{k,l})+ \rho  (c^i_{k-1,l})}{2\delta _{x_1}^2} + \mathcal{O}(\delta _{x_1}^2).  
\\
  	\partial^2 _{x_1,x_2} \rho(c^i_{k,l})&=\frac{\rho(c^i_{k+1,l+1}) - \rho  (c^i_{k+1,l-1})- \rho  (c^i_{k-1,l+1}) + \rho(c^i_{k-1,l-1})}{4\delta _{x_1}\delta _{x_2}} + \mathcal{O}(\delta _{x_1}\delta _{x_2}).  
\end{align}
\end{widetext}
This stencil is completed with the same schemes in the $x_2$-direction. These numerical derivatives can be arranged into matrices so that their multiplication with vectors approximate their respective differentiation. Thus we obtain matrix representations $m_k$ of the operators $\mathcal{B}_k$.

Of course, the schemes presented above only work for the interior boxes of the domain. We have not implemented explicit boundary conditions since the values of the invariant measure of the O-U process at the boundary of the compact domain are almost zero. In any case, boundary conditions should not inject/deplete mass so in other cases where the dynamics populate the boundaries, Neumann boundary conditions are the ones to choose\cite{froyland2013}.

\begin{figure*}
\includegraphics[scale=0.2]{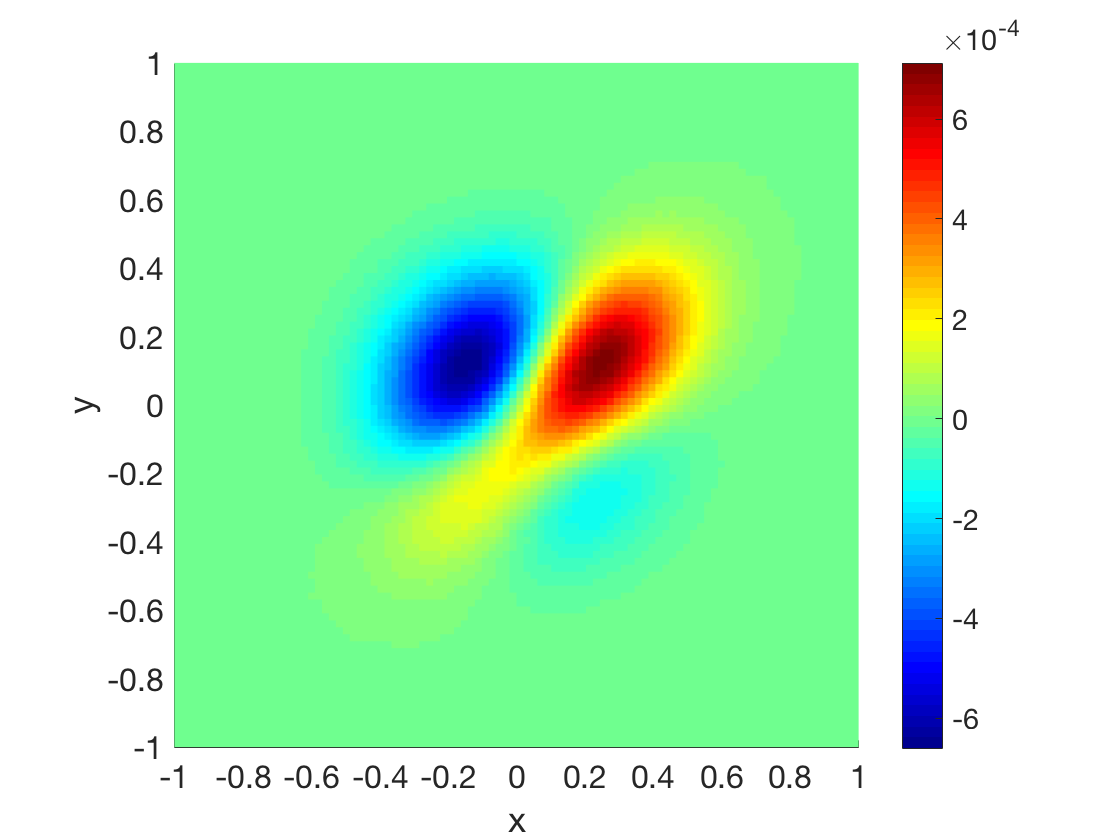}
\includegraphics[scale=0.2]{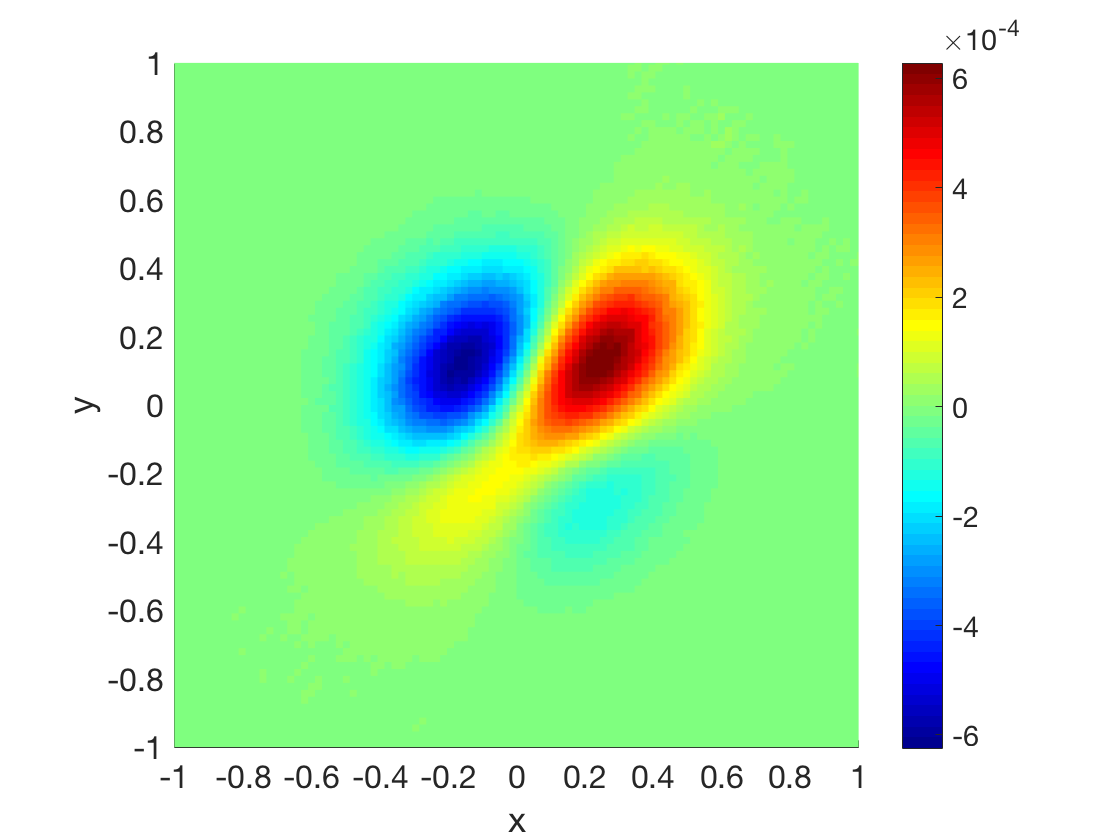}
\caption{\label{total response ou}Response of the O-U to the perturbations considered in Eq.~(\ref{perturbation ou}). The left-hand figure shows the true response calculated by subtracting the unperturbed invariant measure from the perturbed one. The figure on the right is the predicted response calculated from Eq.~(\ref{algebraic expansion 1}).}
\end{figure*}

\begin{figure*}
\includegraphics[scale=0.2]{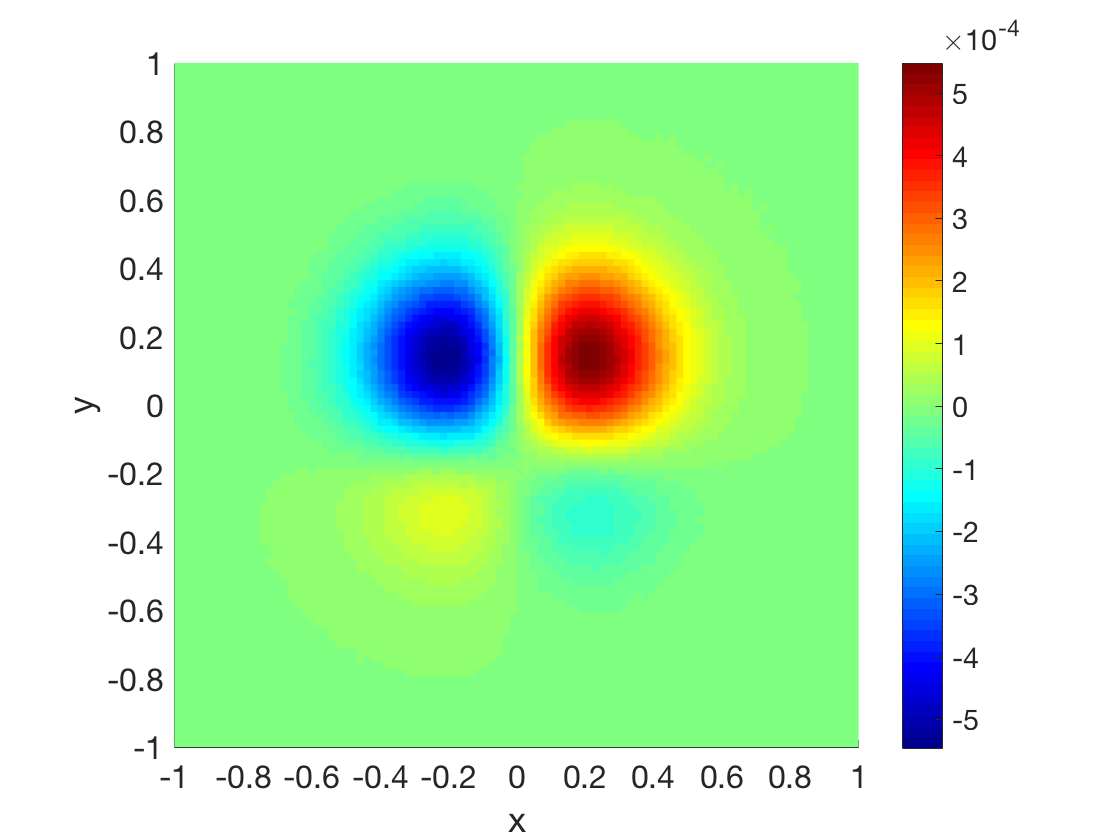}
\includegraphics[scale=0.2]{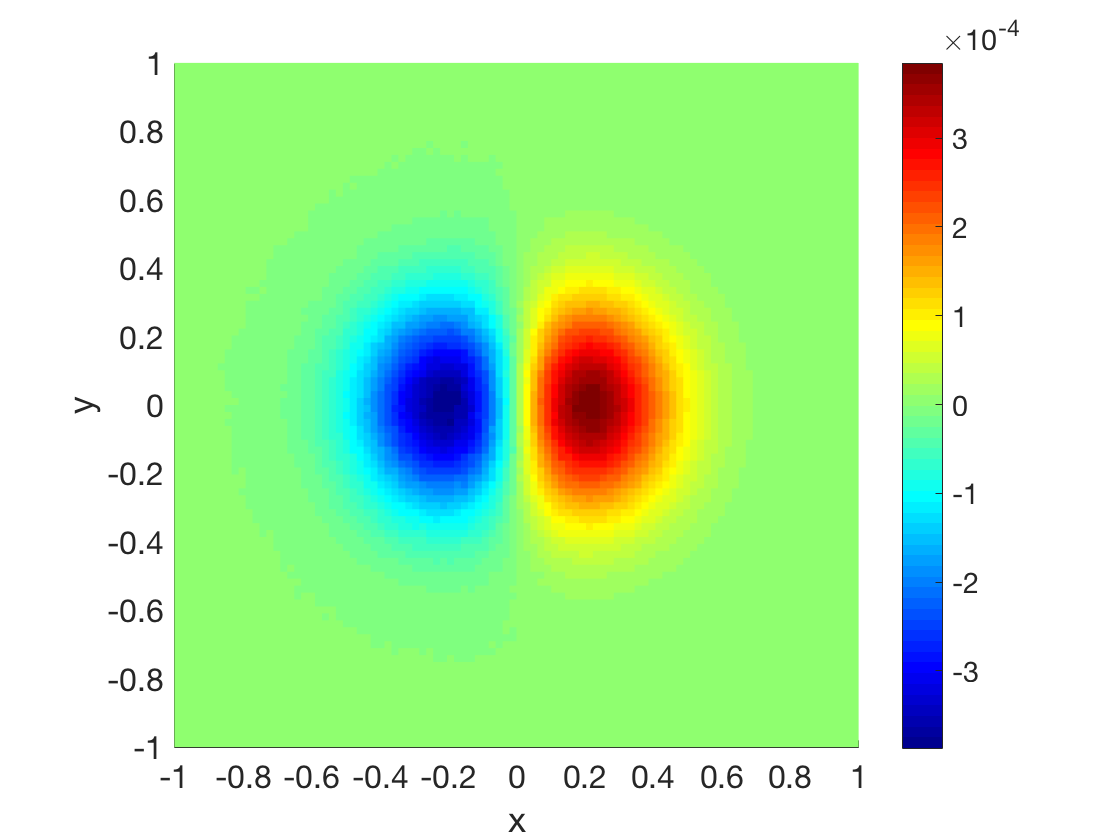}
\includegraphics[scale=0.2]{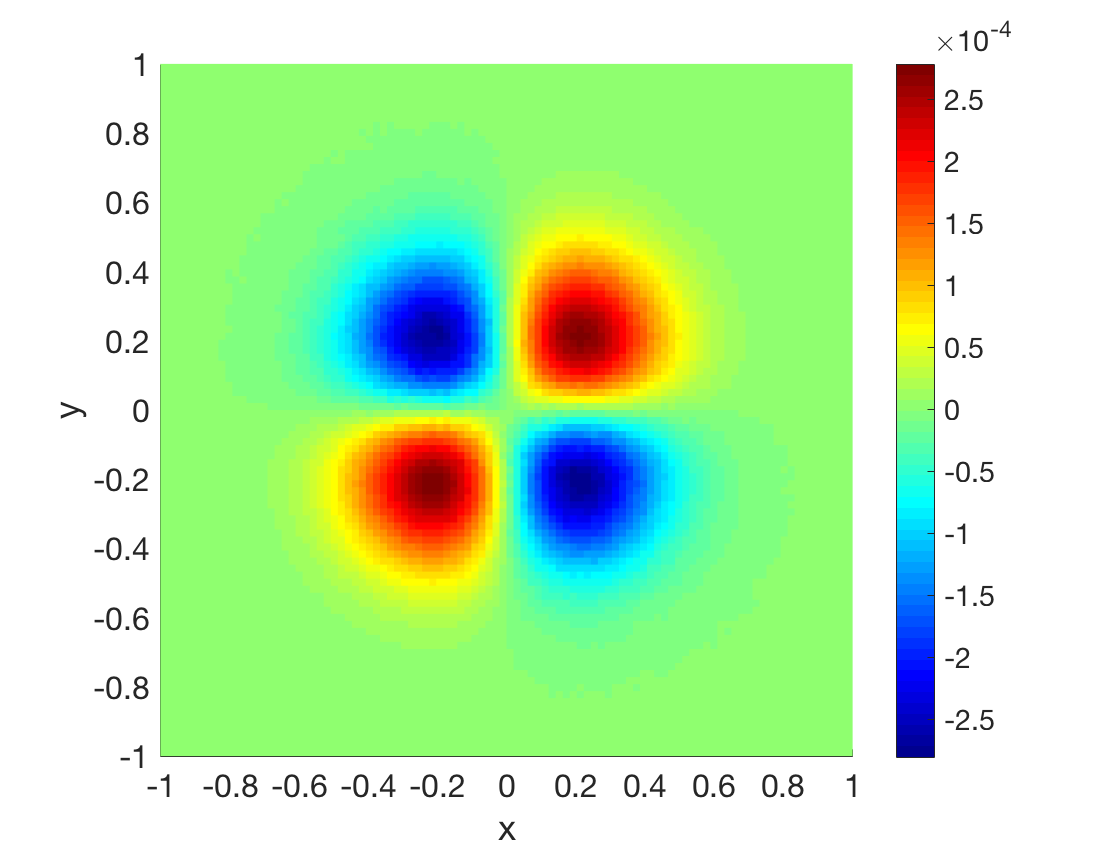}
\includegraphics[scale=0.2]{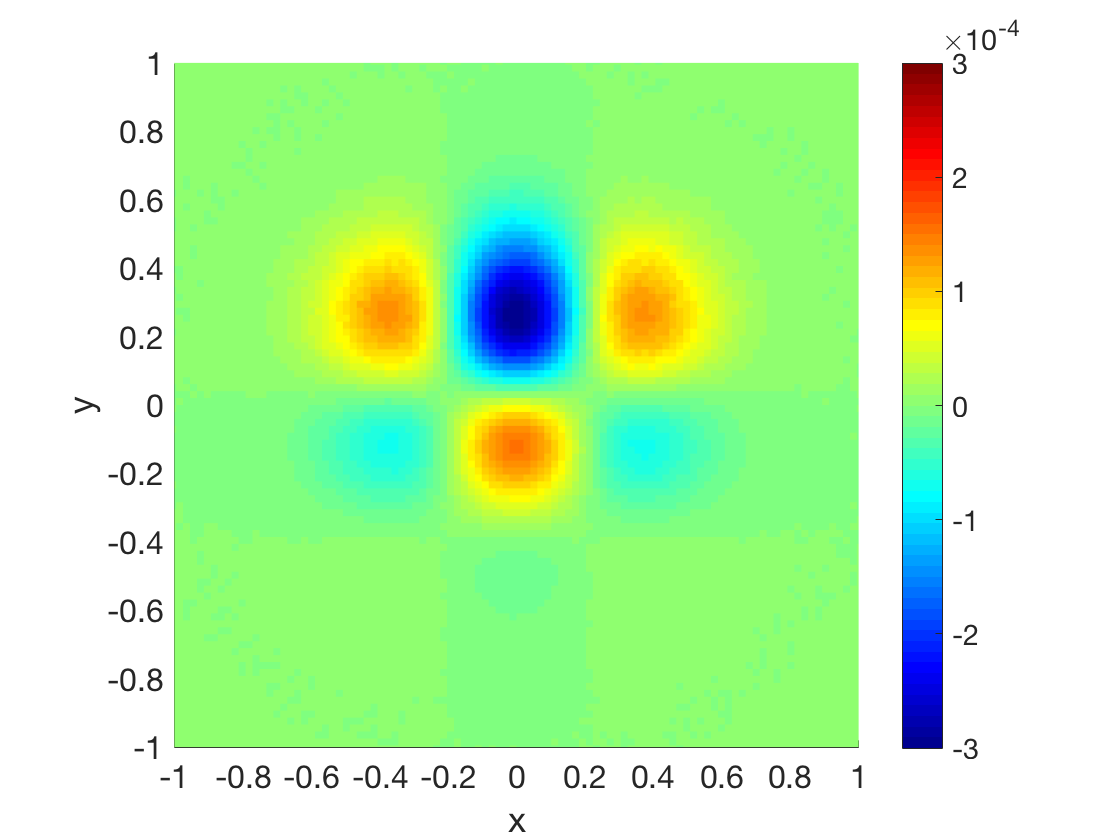}
\caption{\label{response ou plots}The top-left figure shows the linear response of the O-U process with respect to the perturbations considered in Eq.~(\ref{perturbation ou}) calculated by truncating Eq.~(\ref{algebraic expansion 2}) at the first order. The top-right and bottom-left show the linear response to changes in $\epsilon _1$ and $\epsilon _2$ respectively. The bottom-right figure contains the second order correction obtained via Eq.~(\ref{algebraic expansion 2}).}
\end{figure*}

A good compromise was found provided that the resolution was high enough. On Fig.~\ref{total response ou}, we see that the predicted response was well approximated. This is checked on Table~\ref{table ou} where we show that the error of approximating the perturbed invariant measure using the response formulas in Eq.~(\ref{algebraic expansion 2}) presented earlier is small. The linear response (see Fig.~\ref{response ou plots}) is precisely doing what one expects: mass is pumped to the right as a consequence of the mean being shifted and the introduction of correlations inflicts a rotation. Higher order correction terms (see Fig.~\ref{response ou plots}) can also give an insight on how the measure is gradually modified. As a safety check, the sum of the components of the response are checked to add up to (almost) zero, meaning that mass is not introduced or depleted.

\begin{table}
\caption{\label{table ou}The first row refers to the values obtained from integrating the O-U process. The rest of the rows are values obtained via discretisation of the transfer operator. We defined Error1 as the $L^2$ norm of the difference between the coarse-grained perturbed invariant measure and the first order correction. Error2 is the same but with higher order correction terms.}
\begin{ruledtabular}
\begin{tabular}{cccccc}
 &$\langle x \rangle $&$\delta_{\epsilon_1} \left[ x\right]_1$& $\delta_{\epsilon_2} \left[ x\right]_1$ &Error1&Error2 \\ \hline
 O-U & 0 & 1 & 0 & 0 & 0  \\
 $N=10$& $10^{-4}$ &  0.47 &  $3\times 10^{-3}$   &   0.02 & 0.02 \\
 $N=12$& $10^{-4}$  &  0.77 &  $2\times 10^{-3}$  &  7$\times10^{-3}$ & $5\times10^{-3}$ \\
 $N=14$&  $10^{-4}$  & 0.93  & $8\times 10^{-4}$ &  3$\times10^{-3}$ & $8\times10^{-4}$
\end{tabular}
\end{ruledtabular}
\end{table}

\subsection{A Deterministic Dynamical System}


A model of importance in dynamical systems and climatology is the Lorenz 63 system\cite{lorenzdeterministic1963}. Such system is a low-dimensional model of atmospheric convention that, even if it is far from being a realistic model, it possesses chaotic behaviour and exhibits different regimes, just as in the atmosphere. This systems has been the toy-model to illustrate the difficulty of calculating the sensitivity and response in dynamical systems and in the climate system by extension (e.g. Refs.~\onlinecite{leasensitivity2000, wangforward2013}).

From the point of view of the transfer operator, the response of the Lorenz 63 system was computed in Ref.~\onlinecite{Lucarini2016}, by means of applying the perturbation theory for finite-state space Markov chains. However, the perturbed dynamics needed to be sampled. In case of low-dimensional models, this is not a major handicap, but in physically relevant systems this can be an expensive procedure. Therefore, we need more predictive skill.


The Lorenz 63 system is a deterministic dynamical system that is given by the following set of ordinary differential equations:

\begin{equation}\label{lorenz63}
\dot{\mathbf{x}}(t)=\mathbf{F}(\mathbf{x})=\begin{cases}& s (y - x) \\ & x(r - z) - y  \\ & xy -b z  \end{cases},
\end{equation}
for the classical parameter values $s =10$, $b = 8/3$ and control parameter $r =28$. For such parameter values, the Lorenz 63 system displays chaotic dynamics in a singularly hyperbolic attractor that supports an SRB measure\cite{tuckerarigorous2002}. The lack of uniform hyperbolicity implies that the Lorenz 63 system is not Axiom A and therefore one cannot expect response theory to hold. Nevertheless, numerical evidence supports the idea of linear\cite{reicklinear2002} and nonlinear\cite{valerioevidence2009} response to be valid in this system.

The perturbation problem we tackle here is that of changing the value of the control parameter $r \rightarrow r + \epsilon_1$ for $\epsilon_1 \in \mathbb{R}$. This parameter is known as the Rayleigh number and it is proportional to the temperature difference between the convecting layers. The bifurcations associated to this parameter are numerically surveyed in Ref.~\onlinecite{sparrow}. We would also study the additive perturbation on the $z$-variable by adding $\epsilon _2 \in \mathbb{R}$. These perturbations incur a modification on the vector field of the form:
\begin{equation}\label{lorenz63eps}
\dot{\mathbf{x}}(t)=\mathbf{F}(\mathbf{x})+\epsilon_1 \mathbf{G}_1(\mathbf{x})+\epsilon_2 \mathbf{G}_2(\mathbf{x}).
\end{equation}
where $\mathbf{G}_1(\mathbf{x})=[0,x,0]'$ and $\mathbf{G}_2(\mathbf{x})=[0,0,1]'$. Consequently, the Liouville equation changes as in Eq.~(\ref{liouville equation}), with perturbation operators:
\begin{equation}
    \mathcal{B}_1\rho(\mathbf{x},t)=-\nabla \cdot \left( \mathbf{G}_1(\mathbf{x})\rho(\mathbf{x},t) \right) \\
\end{equation}
and
\begin{equation}
    \mathcal{B}_2\rho(\mathbf{x},t)=-\nabla \cdot \left( \mathbf{G}_2(\mathbf{x})\rho(\mathbf{x},t) \right).
\end{equation}

\begin{figure}
\centering
\includegraphics[scale=0.2]{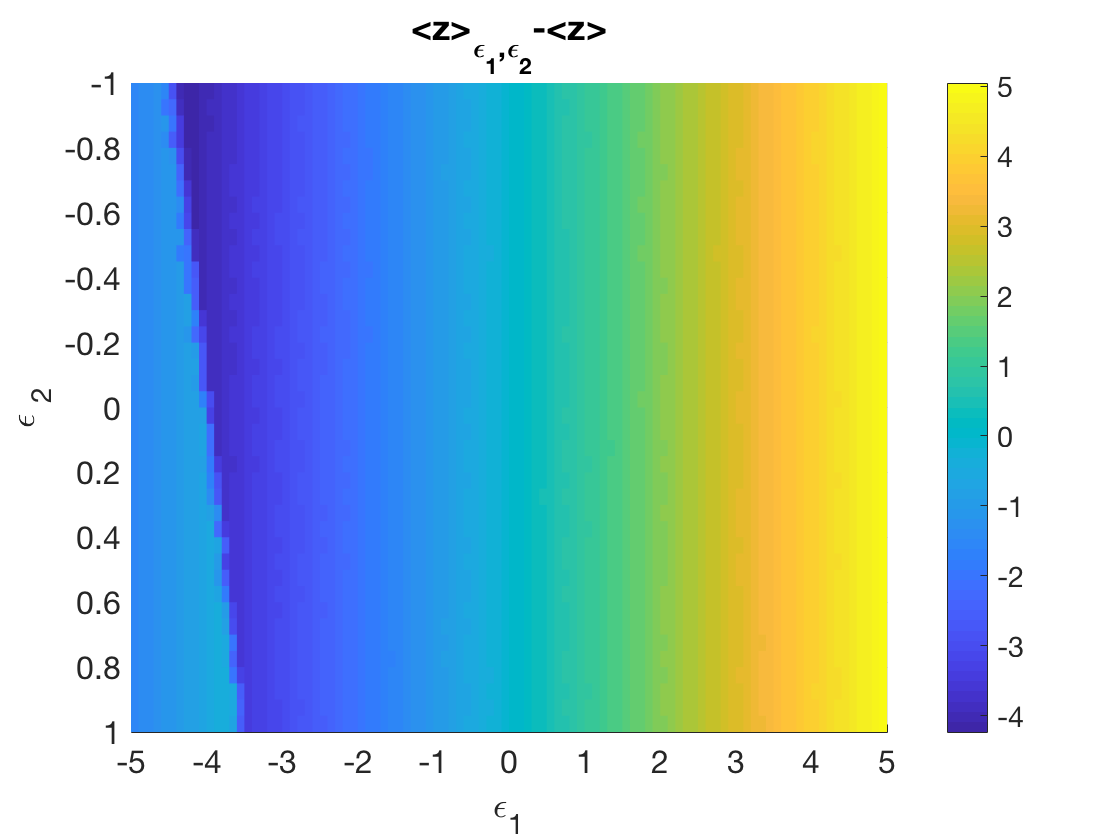}
\caption{\label{reference expectation values} Deviation of perturbed expected values of the observable $z$ for $\epsilon_1 \in [-5,5]$ and $\epsilon_2\in [-1,1]$. The expectation values were computed by integrating the equations for $10^3$ time-units with 20 ensemble members for each value of $\epsilon_1$ and $\epsilon_2$.}
\end{figure}

Introducing forcing provokes a change in the statistics of the system as shown on Fig.~\ref{reference expectation values}, where we considered the example observable $z$ and computed its mean value for equispaced values of $\epsilon_1 \in [-5,5]$ and $\epsilon_2\in [-1,1]$. When $\epsilon_1 \approx -4$, a bifurcation is traversed producing a non-smooth change in the mean value of the observable. Far away from this bifurcation point, the statistics changes smoothly with respect to $\epsilon _1$ and $\epsilon _2$. For the calculation of these means, long integrations were considered so that the outcome is uniquely determined. This is possible thanks to the existence of an SRB measure\cite{leasensitivity2000}.

Notice that differentiating Eq.~(\ref{perturbed fkp equation}) with respect to $\epsilon _k$ ($k=1,2$) gives $\mathcal{B}_k$. So a way of obtaining a matrix representation of $\mathcal{B}_k$ is
\begin{equation}\label{increment}
    \epsilon_k m_k = \mathcal{M}_{\epsilon_k}^{dt} - \mathcal{M}^{dt}.
\end{equation}
Where $\mathcal{M}_{\epsilon_k}^{dt}$ is the transition matrix empirically constructed from the perturbed equations. This approach was followed previously\cite{Lucarini2016} and served to calculate the linear response of the Lorenz 63 system, however, it involves two (or more) long integrations of the equations, something we want to avoid. We explain the methodology carried out here in the next section.

\subsubsection*{Numerics: transition and perturbation matrices}

The invariant measure of the Lorenz 63 system is supported on attractor, which by definition is compact. Since we know that the Lorenz attractor is contained in an absorbing closed ellipsoid on phase-space\cite{sparrow}, we are certain that the square defined by the Cartesian product $[-20,20]\times [-30,30] \times [0, 50]$ will contain the attractor. Then, to obtain the box partition, we divide into two each axis and repeat the procedure in each resulting segment. This way, a total of $2^N$ boxes $\{B_i\}_{i=1}^{2^N}$ were constructed for the values of $N=12,15$ and $18$. As mentioned earlier, the box discretisation was carried out using GAIO\cite{gaio}.

\begin{figure*}
\includegraphics[scale=0.2]{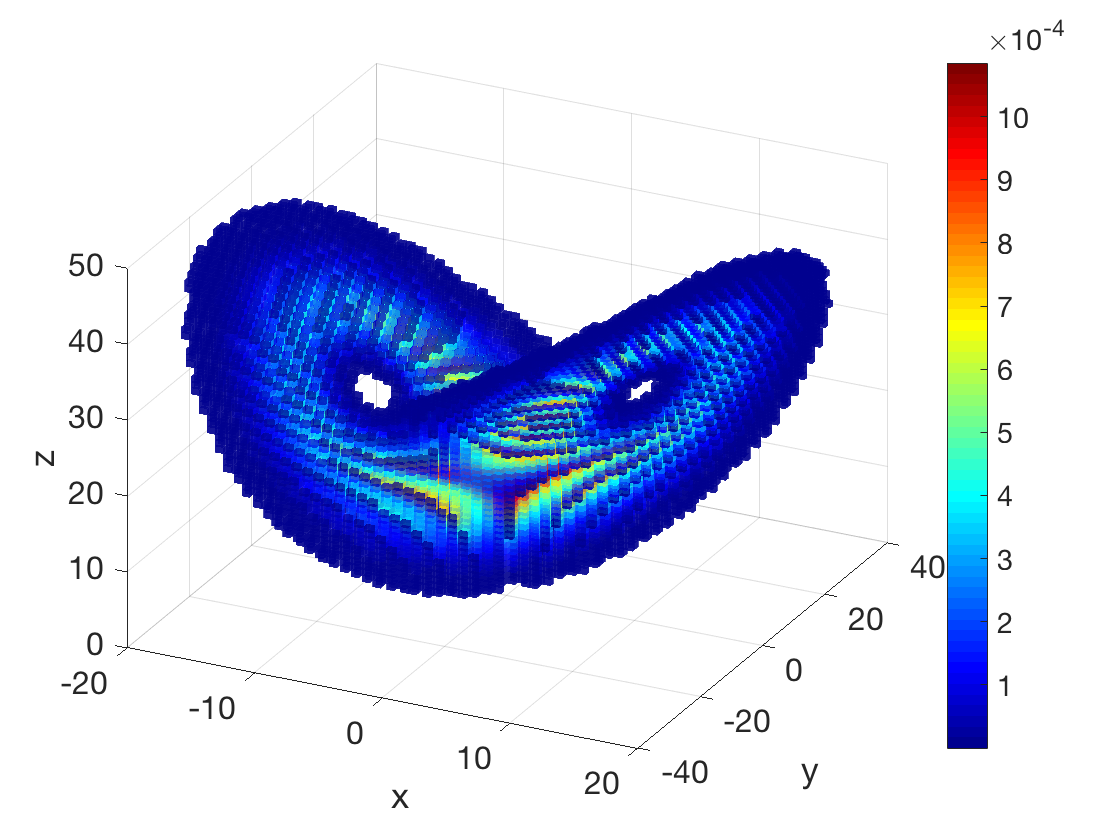}
\includegraphics[scale=0.2]{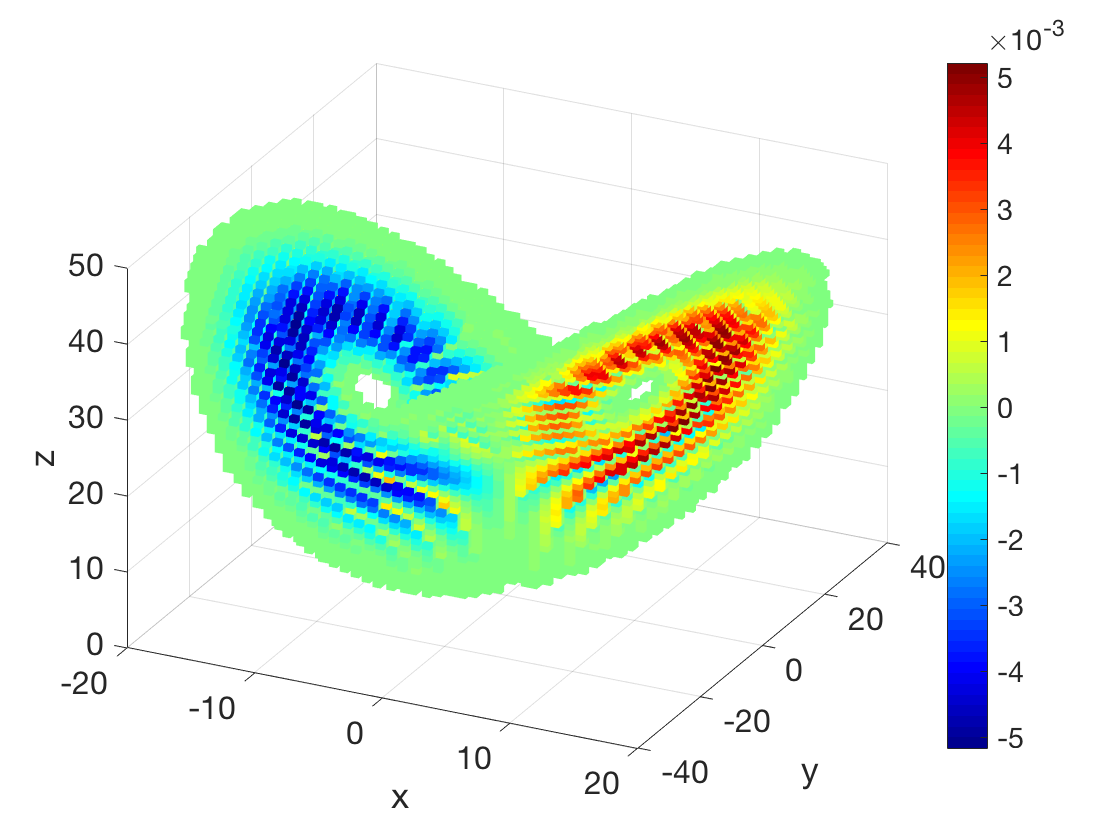}
\caption{\label{invariant measure}Coarse-grained Lorenz 63 invariant measure (left) and the discretised operator $\mathcal{B}_1$ applied onto the invariant measure (right).}
\end{figure*}

To obtain the sample points of the dynamics, we integrated the model for $10^5$ time-units with a time-step of $dt=10^{-3}$ time-units. After a spinup of ten percent of the points, we count the transitions between the boxes with a time-lag of $dt$ time-units. This gives the transition matrix $\mathcal{M}^{dt}$. Thus, a coarse-grained estimation of the invariant measure can be readily obtained by solving the eigenvalue problem in Eq.~(\ref{eigenvalue problem}) which is plotted on Fig.~\ref{invariant measure}.

Regarding the perturbation operators, the same techniques as in the O-U process where employed. The difference here is that we are dealing with an invariant measure that is singular with respect to the Lebesgue measure and therefore is supported in a complicated set, as illustrated by Fig.~\ref{invariant measure}. Hence, one needs to take care of the boundary boxes by simply considering forward/backward approximations of the derivatives at those boxes. Thus was estimated $\mathcal{B}_1$ applied to the invariant measure, depicted on the right of Fig.~\ref{invariant measure}.

\begin{figure*}
\includegraphics[scale=0.2]{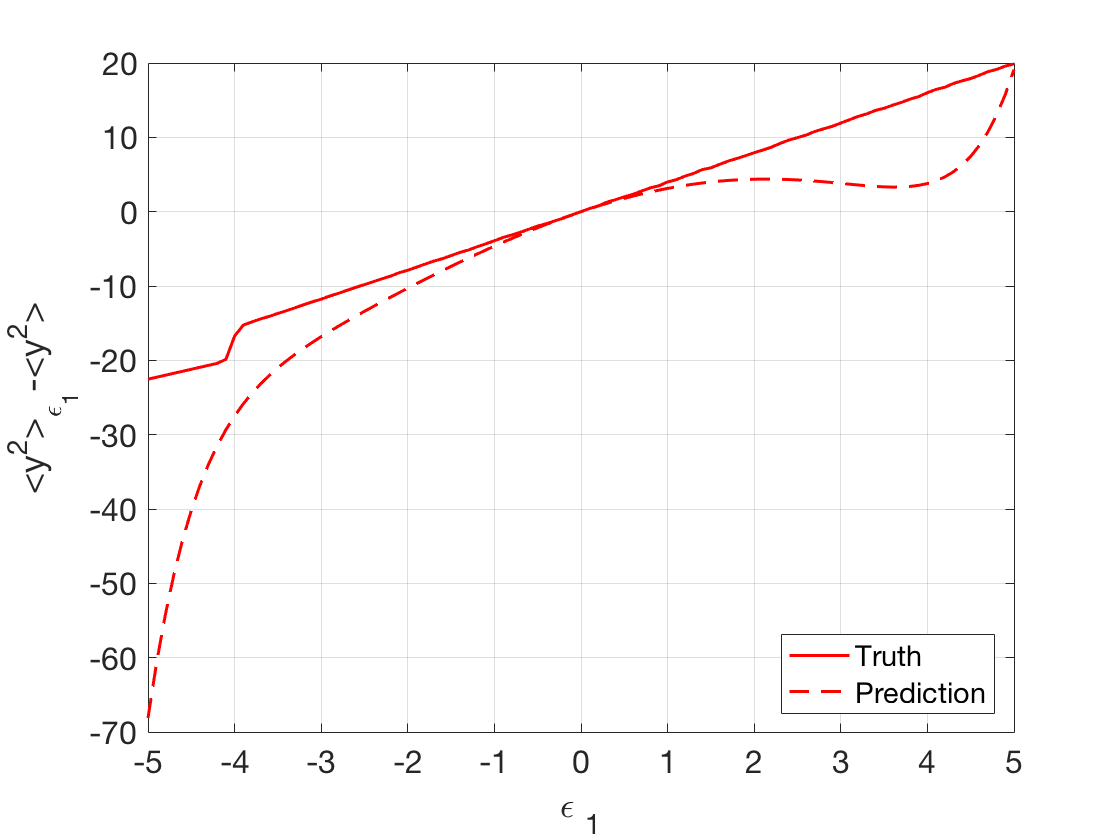}
\includegraphics[scale=0.2]{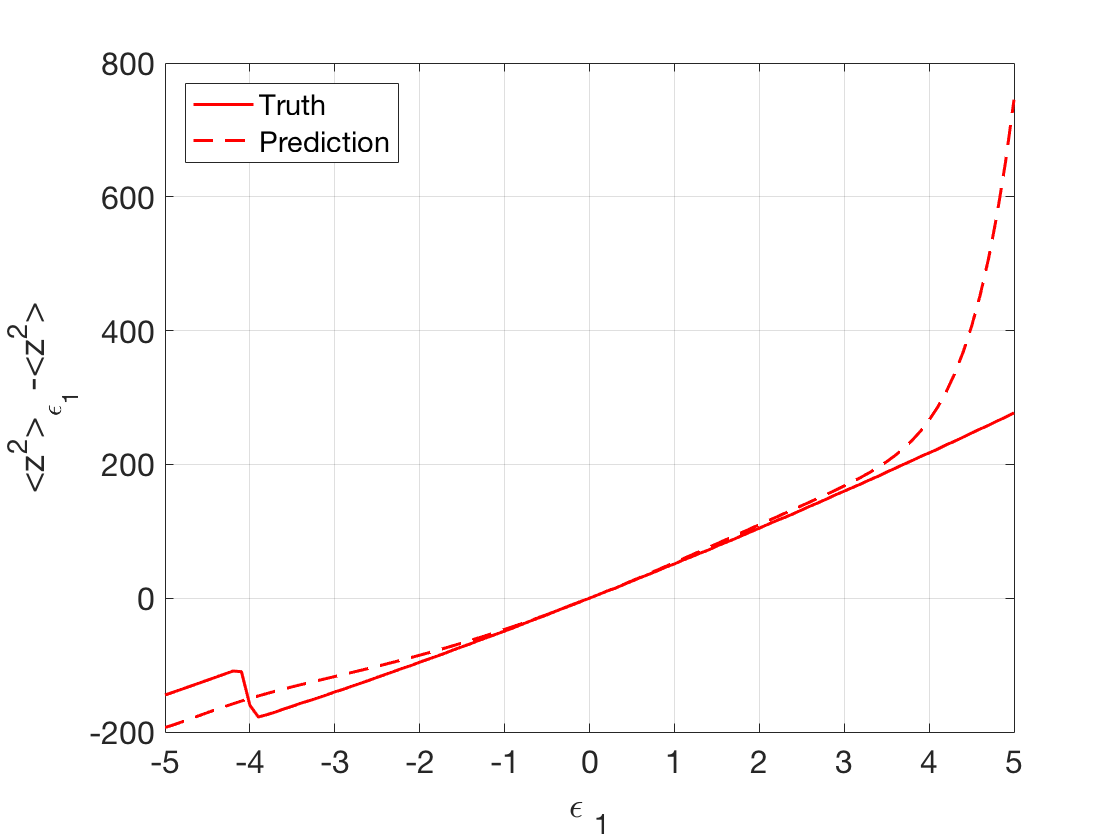}
\includegraphics[scale=0.2]{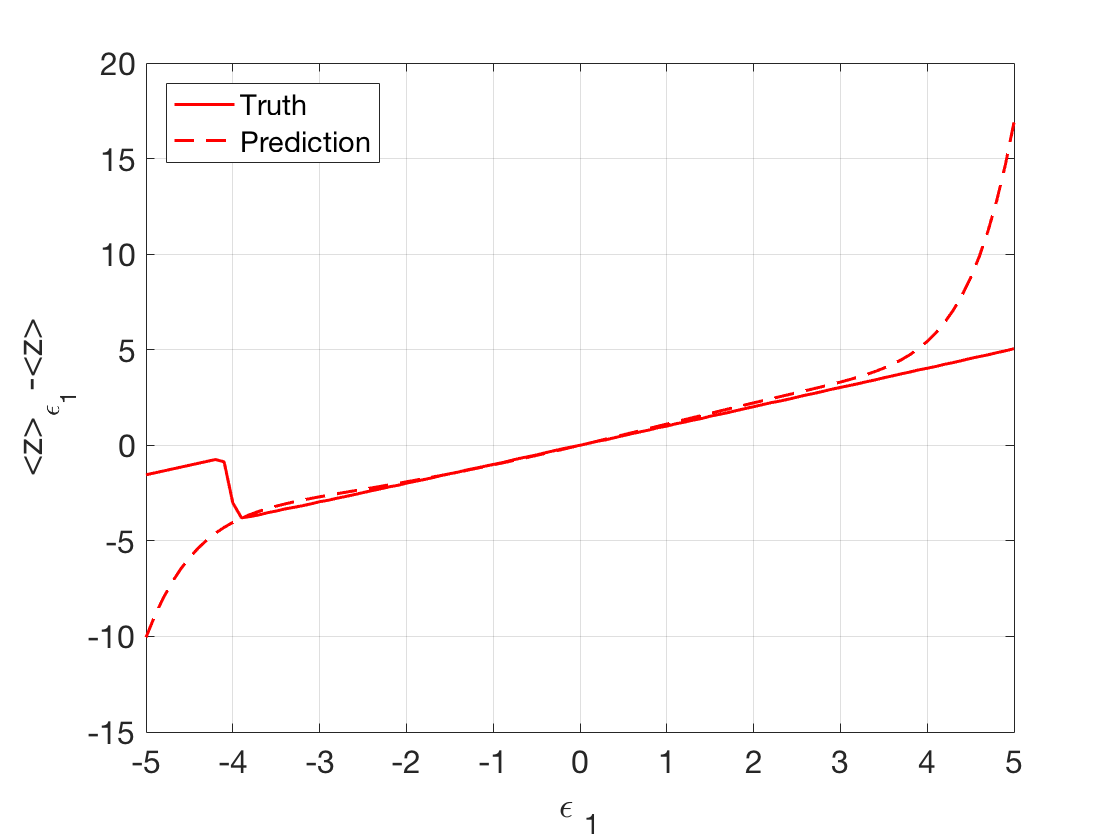}
\includegraphics[scale=0.2]{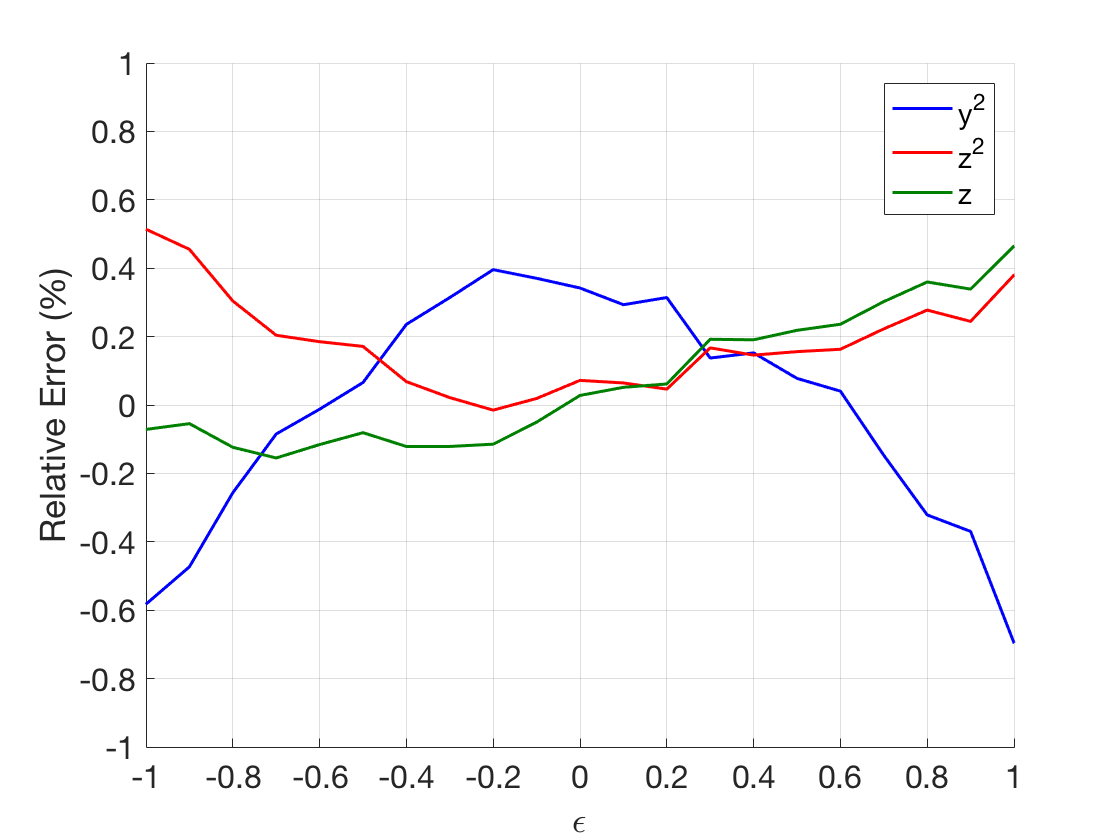}
\caption{\label{predictions l63}Predictions for perturbations on $r$. Mean values of the observables indicated in the panels are predicted using Eq.~($\ref{algebraic expansion 2}$) and plotted against the true means observed empirically via integrations of the system.}
\end{figure*}

\begin{figure*}
\includegraphics[scale=0.2]{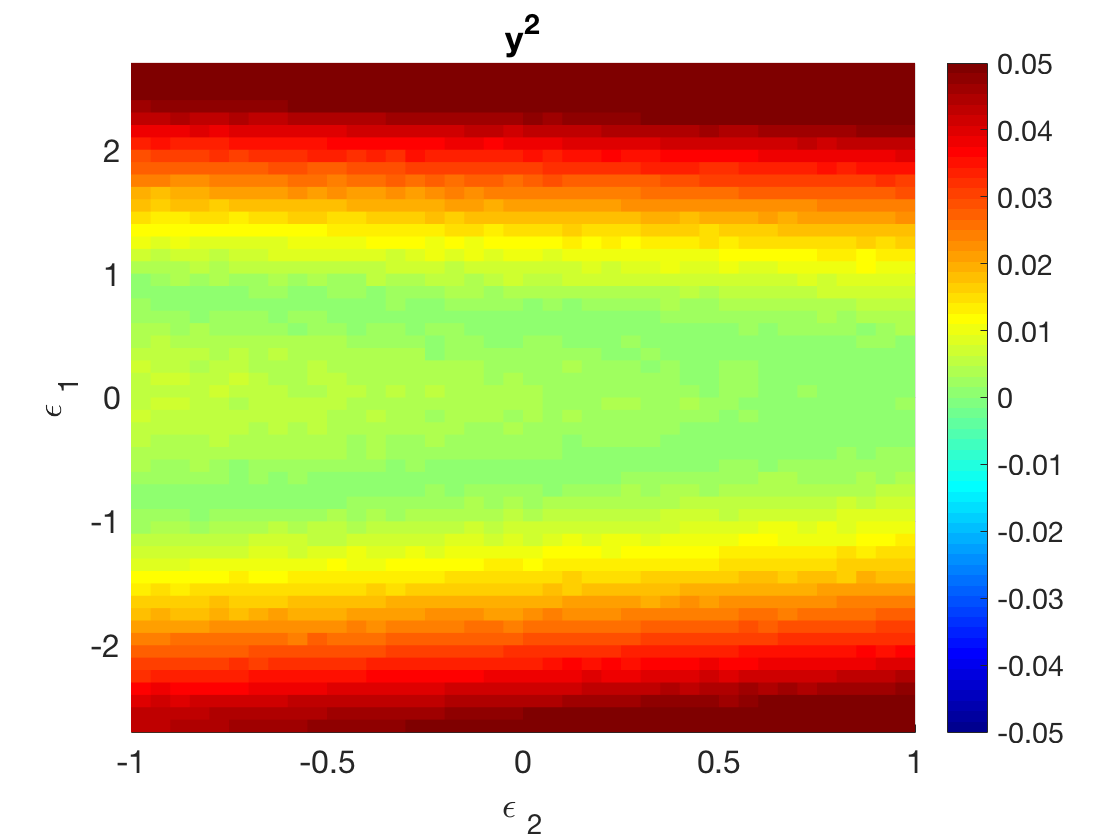}
\includegraphics[scale=0.2]{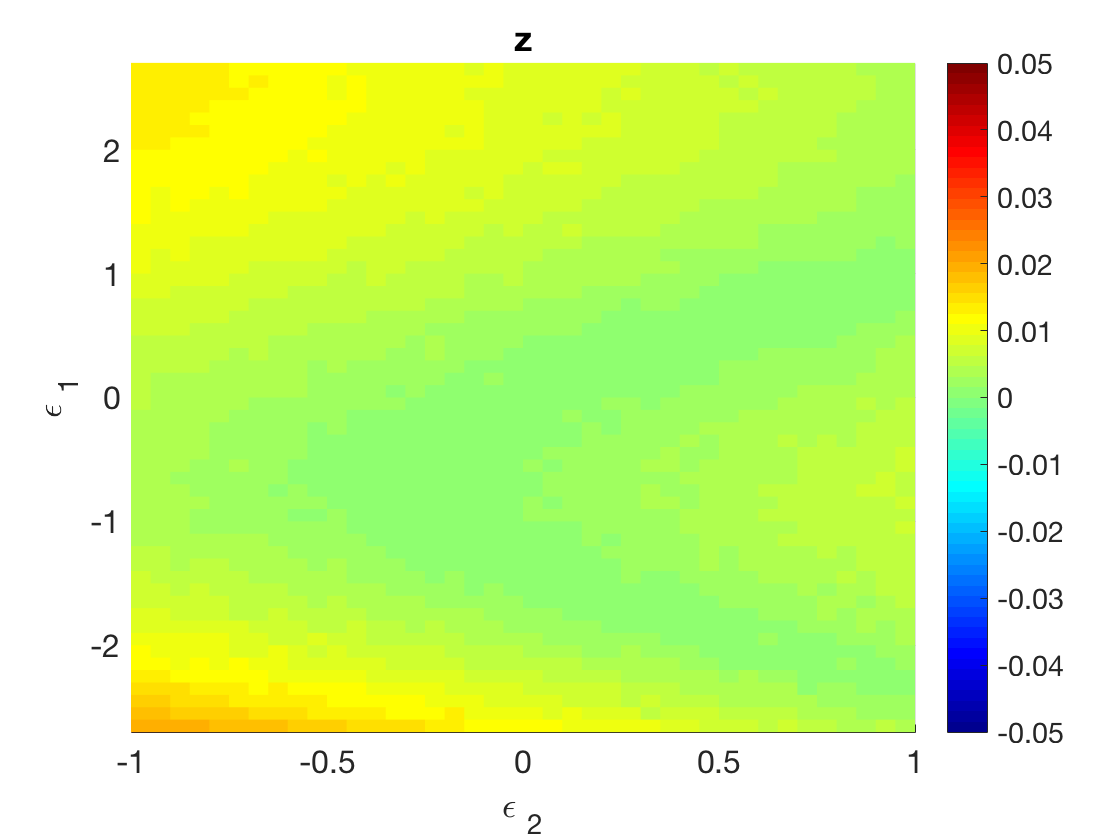}
\caption{\label{2d prediction} Relative error $(\%)$ of the prediction of the expected values of the observables indicated in the plots. The prediction was done by evaluating Eq.~(\ref{algebraic expansion 2}) and the true values were obtained empirically by integrating the equations for each value of $\epsilon _1$ and $\epsilon _2$.}
\end{figure*}

The results of applying this methodology are presented in Fig.~\ref{predictions l63}, where we show that Eq. (\ref{algebraic expansion 2}) can indeed approximate the expectation values certain observables for a wide range of values of $\epsilon _1$. We underline that these plots demonstrate the validity of the formulae not only to compute the linear response but to predict the statistics of the system. As is natural, the formulas cannot be expected to work for large values of the perturbation parameter, let alone beyond the bifurcation point. When considering the simultaneous forcings $\epsilon _1\mathbf{G}_1$ and $\epsilon _2 \mathbf{G}_2$, we examine the efficiency of the formulas (Table~\ref{table l63}). We calculate the expectation value of certain observables and check that we can approximate their mean for small values of $\epsilon _1$ and $\epsilon_2$. Also, the linear response is calculated. As we see in the last row, refining the resolution does not always improve the results. This is due to the fact that the length of the integration has to be severely extended in order to sample the boxes covering the domain. In general, the skill of the methodology is shown in Fig.~\ref{2d prediction} where the expectation values of some observables are predicted using the response formula Eq.~(\ref{algebraic expansion 2}). Of course, the validity was only tested within the convergence interval.

\begin{table*}
\caption{\label{table l63}Expectation values with respect to the unperturbed and perturbed ($\epsilon_1= 0.1$ and $\epsilon _2 =0.1$) invariant measure. In the first row time averages were used whereas the rest indicate the expection values obtained by means of evaluating Eq.~(\ref{algebraic expansion 1}) with a transition matrix of size $N\times N$.}
\begin{ruledtabular}
\begin{tabular}{ccccccccc}
 &$\langle y^2\rangle $&$\langle z \rangle$&$\langle y^2 \rangle_{\epsilon_1,\epsilon _2}$& $\langle z \rangle_{\epsilon_1,\epsilon _2}$ &$\delta_{\epsilon_1} \left[ y^2\right]_1$&$\delta _{\epsilon_1} \left[ z\right]_1$&$\delta _{\epsilon_2} \left[ y^2\right]_1$& $\delta _{\epsilon_2}\left[ z\right]_1$ \\ \hline
 L-63 & 81.20 & 23.54&81.44&23.65 &3.95 &1.01& -1.47 & 0\\
 $N=12$&82.37 &  23.55 &  82.50   &   23.62&3.01   &  0.89 &  -1.65  &  -0.14 \\
 $N=15$& 81.48  &  23.55 &  81.70   &  23.65 & 3.97  &  1.08 &  -1.62  &  -0.11\\
 $N=18$& 81.25   & 23.55  & 81.30 &  23.65 & 2.68  & 1.11  & -1.57   &  -0.09

\end{tabular}
\end{ruledtabular}
\end{table*}

\pagebreak
\section{Discussion	}

In this work, we take the point of view of statistics to understand the effect of perturbations on dynamical systems. Under this setting, the transfer operator is the essential object of study. By means of projecting such operator onto a finite dimensional vector space, it is possible to obtain empirically constructed matrix representations of the transfer operator via Markov modelling. Further, the properties of these matrices allow us to, up to finite precision, describe the dynamical system of interest. 

Exploiting the stochastic (or Markovian) structure of the projected transfer operator, we consider multiple perturbations and express the resulting perturbed invariant measure as a series expansion describing the response at all orders of nonlinearity. Alongside previous work\cite{schweitzerperturbation1968,Lucarini2016, froylandoptimal2018}, it is possible to link the probabilistic concepts of finite Markov chains to dynamical systems. Namely, the mixing rate of the unperturbed chain indicated by its second eigenvalue or more generally its ergodicity coefficient determine the validity of the perturbative expansions. This agrees with more theoretical considerations with the so-called \emph{spectral gap condition}\cite{Hairer2010}, key to establish linear response.

The linear component in the perturbative expansion is the so called linear response and it is of physical interest\cite{marconireview2008}. This quantity is accessed by constructing suitable response operators and indicates the sensitivity of a dynamical system to prescribed perturbations. Here we emphasize the need of gaining predictive skill: given a perturbation to the vector field (possibly induced by tuning several paramters), can we calculate the sensitivity of the system \emph{a priori}? For such purpose, we examined the Fokker-Planck/Liouville equation to identify the operators that incur the perturbations on the evolution of densities. Then, by considering simple finite-difference methods, we were able to model (to finite-precision) matrix perturbations that allowed to exploit the perturbative formulas giving us access to the linear and nonlinear response of the systems. Notice that using this method, we only need one integration of the (unforced) model to determine the response and sensitivity.

The two numerical experiments performed were intrinsically different. The Ornstein-Uhlenbeck process is a stochastic dynamical system with an invariant measure with a density (with respect to the Lebesgue measure) which is smooth. Therefore, differential operators (discretised using finite-differences) should work correctly. On the other hand, the Lorenz 63 model is a dissipative deterministic system with an attractor with Lebesgue-zero measure that is singularly hyperbolic. This translates into the fact that linear response theory is not theoretically proved\cite{ruelle2009}. Moreover, the invariant measure of such system is not smooth, so the usual notion of differentiation does not hold. In the experiments, we coarse-grained the invariant measure to obtain a vector estimate. The corresponding differential operators were applied giving a good compromise, suggesting that at a coarse-grained level, the usual methods for numerical differentiation should work. Previous investigations\cite{froyland2013} illustrate that coarse-graining the domain inherently provokes the introduction numerical diffusion thus smoothening the invariant measure.

The partitioning of phase-space is an arbitrary decision of the modeller. In these numerical experiments, the partitioning is considered to be uniform: boxes are of equal size. This is not optimal, since there might be regions of phase-space that need more refining than others\cite{dellnitzon1999,froylandex2001}. The reason for choosing equally sized boxes is that the invariant measures obtained gave a good compromise when approximating the expectation value of observables. In any case, comparing box discretisations is not the target of the paper.

The low dimensionality of the problems considered in this paper allows to perform the box subdivision on the whole phase-space. Unfortunately, many physically relevant models posses a domain which is high-dimensional not to say infinite-dimensional. In these cases, it appears intractable to work on the complete domain. Because of this dimensionality barrier, reduced phase-spaces are considered. Results along this line support the applicability of the transfer operator methods in a climatological context, see, e.g., Refs.~\onlinecite{chekrounrough2014, Tantet2015, tantetcrisis2018}. However, the inherent loss of Markovianity in the dimensionality reduction requires control on the memory effects introduced by the hidden variables (see, e.g., Ref.~\onlinecite{kondrashovdata2015}), something that complicates the study of the response, as pointed out in Ref.~\onlinecite{lucariniwouters} where the robustness of reduced systems with the presence of forcing is assessed. Further research should be oriented on adapting this methodology based on the transfer operator in higher-dimensional systems with vistas to assessing and predicting the sensitivity of physically relevant systems.

%

\nocite{*}
\bibliography{phd1_mpp.bib}

\end{document}